\documentclass[twocolumn,pre]{revtex4}   

\usepackage{dcolumn}
\usepackage{amsmath}
\usepackage{bbm}

\usepackage{graphicx}
\usepackage{bm}

\newcommand{\defn}{\textit}
\newcommand{\mat}{\mathbf}
\renewcommand{\vec}{\mathbf}
\newcommand{\e}{\mathrm{e}}
\newcommand{\ii}{\mathrm{i}}
\newcommand{\Tr}{\mathop\mathrm{Tr}}
\renewcommand{\Im}{\mathop\mathrm{Im}}
\newcommand{\etal}{{\it{}et~al.}}
\newcommand{\Ord}{\mathrm{O}}
\newcommand{\from}{\leftarrow}
\newcommand{\av}[1]{\langle#1\rangle}
\newcommand{\set}[1]{\lbrace#1\rbrace}
\newcommand{\oin}{\omega_{\textrm{in}}}
\newcommand{\oout}{\omega_{\textrm{out}}}
\newcommand\cin{c_\textrm{in}}
\newcommand\cout{c_\textrm{out}}

\begin{document}

\title{Message passing methods on complex networks}
\author{M. E. J. Newman}

\affiliation{Department of Physics and Center for the Study of Complex Systems,\\University of Michigan, Ann Arbor, MI 48109, USA}

\begin{abstract}
Networks and network computations have become a primary mathematical tool for analyzing the structure of many kinds of complex systems, ranging from the Internet and transportation networks to biochemical interactions and social networks.  A common task in network analysis is the calculation of quantities that reside on the nodes of a network, such as centrality measures, probabilities, or model states.  In this review article we discuss message passing methods, a family of techniques for performing such calculations, based on the propagation of information between the nodes of a network.  We introduce the message passing approach with a series of examples, give some illustrative applications and results, and discuss the deep connections between message passing and phase transitions in networks.  We also point out some limitations of the message passing approach and describe some recently-introduced methods that address these limitations.
\end{abstract}

\maketitle

\section{Introduction}
\begin{figure}[b]
\begin{center}
\includegraphics[width=\columnwidth]{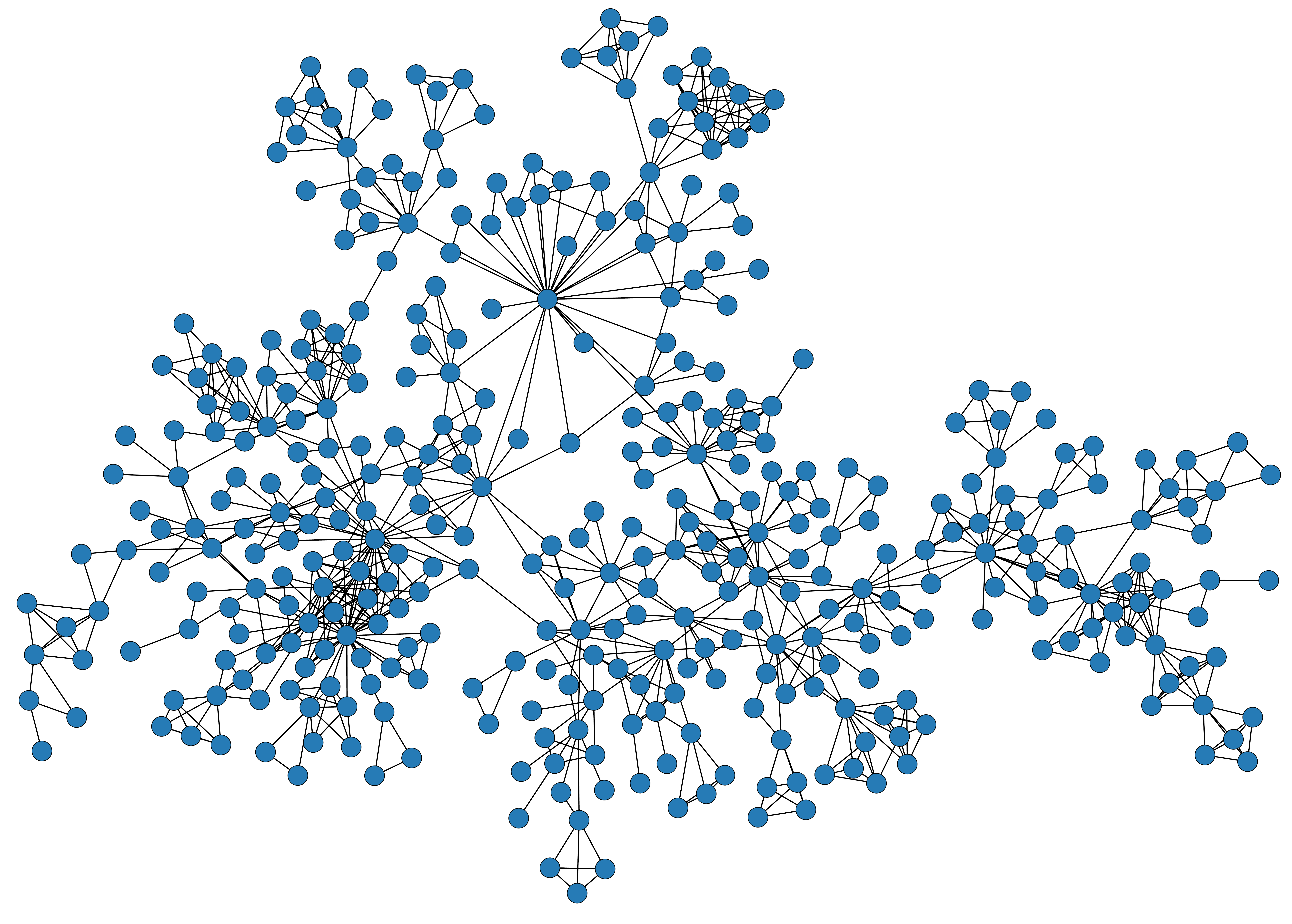}
\end{center}
\caption{A small social network representing the pattern of collaborations among a group of 379 scientists.}
\label{fig:collabs}
\end{figure}

A network, for the purposes of this paper, is a set of points joined together in pairs by lines, as shown in Fig.~\ref{fig:collabs}.  In the nomenclature of the field the points are called \defn{nodes} or \defn{vertices} and the lines are called \defn{edges}.  Networks provide a convenient tool for representing the pattern of connections or interactions within complex systems of many types.  The Internet, for example, is a network of computers joined by data communications.  The Web is a network of pages connected by hyperlinks.  Biochemical networks, such as metabolic networks and protein interaction networks, are networks of chemicals and their interactions.  And there is a long history of the quantitative study of social networks, which means not only modern online networks like Facebook and Twitter, but also off\-line social networks such as networks of friendship, collaboration, and acquaintance, and the contact networks over which diseases spread.  Figure~\ref{fig:collabs}, for instance, shows the patterns of collaboration among a group of scientists.  In recent decades a large body of knowledge has built up concerning methods for analyzing and interpreting network data, and databases have been assembled containing data sets and measurements for thousands of different networks.  For an introduction to this vibrant and fascinating field see Refs.~\cite{Strogatz01,Watts03,Boccaletti06,Barabasi16,Newman18c}.

Development of the theory and practice of network analysis has been a multidisciplinary undertaking spanning mathematics, computer science, physics, statistics, the social sciences, and other fields and has incorporated techniques from many areas, including linear algebra and spectral theory, probability and combinatorics, randomized models, statistics and machine learning, random matrix theory, and a wide range of numerical methods.  In this paper we focus on a less well known but powerful class of methods variously called \defn{message passing}, \defn{belief propagation}, or \defn{cavity methods}.  These methods have theoretical foundations going back to the 1930s~\cite{Bethe35} but in their modern incarnation they were first formulated by Pearl in the 1980s~\cite{Pearl82}.  They have been widely applied to problems in computer science and, less commonly, statistical physics and other fields~\cite{MM09,MPZ02,YFW03}.  In the network applications we consider here the first goal of message passing methods is to calculate properties of individual nodes in networks, such as the state they are in at a given time, but message passing can also be applied to the calculation of global network properties and can serve as a starting point for further analyses, for example of structural phase transitions in networks.

We begin in the following section by introducing message passing on networks with some simple examples, then in Section~\ref{sec:phase} we show how the message passing equations for a network define a dynamical system whose fixed points and bifurcations correspond to phase transitions in the network.  In Section~\ref{sec:applications} we give a number of further examples of the method, illustrating its use both as a computational tool and as a doorway to understanding phase transitions.  In Section~\ref{sec:loopy} we discuss a fundamental limitation of traditional message passing methods---that they are exact only on locally loop-free networks---and we describe a recently introduced approach that sidesteps this limitation, allowing message passing to be applied to a wide variety of networks, no matter what their structure.  In Section~\ref{sec:concs} we offer our conclusions.

\section{Message passing}
\label{sec:mp}
\begin{figure}
\begin{center}
\includegraphics[width=5.5cm]{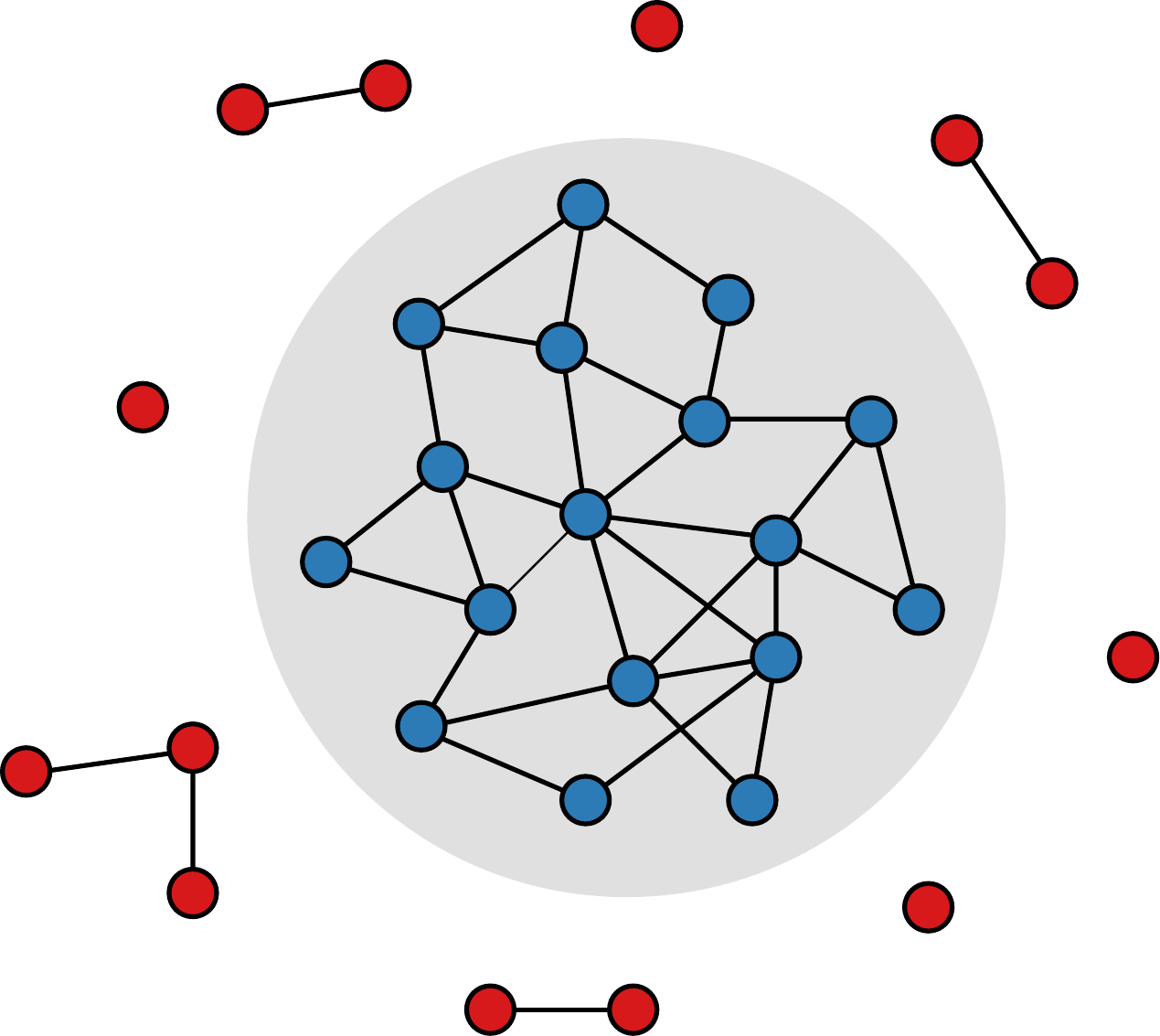}
\end{center}
\caption{Most networks consist of one large group or giant component of connected nodes (circled) and multiple small components or unconnected single nodes.}
\label{fig:structure}
\end{figure}

We begin our discussion with a simple example to demonstrate the message passing method.  Figure~\ref{fig:structure} illustrates a prominent feature of almost all networks, the so-called \defn{giant component}.  In most networks a significant fraction of the nodes are connected to form a large contiguous cluster or giant component, as shown in the center of the figure, while the remainder of the nodes are grouped into a number of small components, shown around the edges.  Networks do not normally contain more than one giant component because the chances of two such components existing yet not being connected together is vanishingly slim.  At the same time a network with no giant component is, for most practical purposes, not really a network at all and so would not be studied in the first place.  As a result, almost every network of scientific or technological importance has the qualitative structure shown in Fig.~\ref{fig:structure}.

Let us consider the following question: given a complete network of $n$ nodes and a specified node within that network, does the node in question belong to the giant component?  To be more precise, we will define the giant component to be the largest component of connected nodes in the network (a slight abuse of nomenclature, but one that will cause us no problems here).  There exist simple computer algorithms that will traverse a network and construct all of its components, allowing us to answer our question quickly, but here, for illustrative purposes, we will take a different approach.

Let the $n$ nodes in our network be labeled by $i = 1\ldots n$ in any convenient order and let $\mu_i$ be the probability that node~$i$ does \emph{not} belong to the giant component.  This is a somewhat trivial example, since $\mu_i$ is always either 0 or~1---either the node belongs to the giant component or it doesn't---but it is a useful example nonetheless and we will soon get to more complex ones.

The crucial point to notice is that node~$i$ is not in the giant component if and only if none of its network neighbors are in the giant component.  If even one of its neighbors belongs to the giant component then $i$ also belongs.  Hence we might write
\begin{equation}
\mu_i = \prod_{j\in \mathcal{N}_i} \mu_j,
\label{eq:wrong}
\end{equation}
where $\mathcal{N}_i$ is the set of nodes that are neighbors of~$i$.  This equation says the probability that $i$ is not in the giant component is equal to the probability that none of its neighbors are.

Trivial though it seems, however, this equation is wrong for two reasons, one obvious and one more subtle.  The first and more obvious reason is that it assumes independence of the neighbors.  In multiplying the probabilities~$\mu_j$ we are assuming that the membership of one neighbor in the giant component is independent of the membership of another.  This is violated if, for instance, two neighbors are connected by an edge, in which case either both are in the giant component or neither is, so their probabilities~$\mu_j$ are correlated.  Normally when applying message passing methods one assumes independence, which may be correct, at least approximately, in some cases but definitely is not in others.  We will, for the moment, make the same assumption of independence here, but in Section~\ref{sec:loopy} we will show how to relax it and apply message passing to networks where the states of nodes are correlated.

\begin{figure}
\begin{center}
\includegraphics[width=6.5cm]{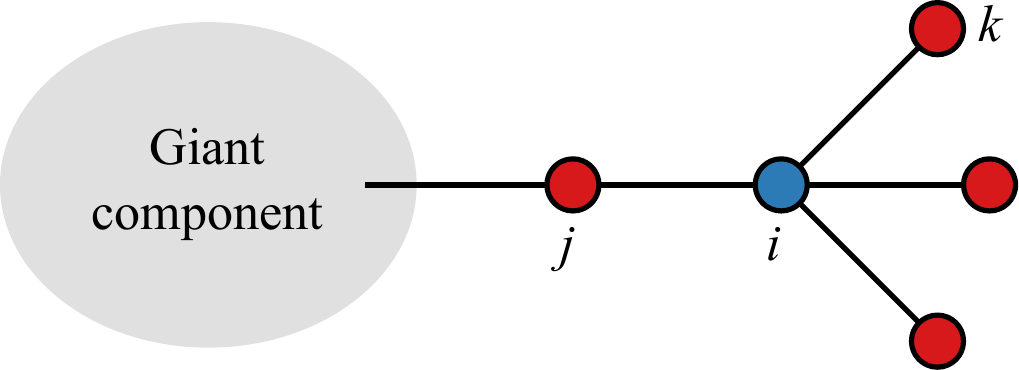}
\end{center}
\caption{Node $i$ is connected to the giant component via its neighbor~$j$.  Its other neighbor~$k$ is also connected to the giant component, but only via~$i$, so $k$ cannot be responsible for $i$'s connection to the giant component.}
\label{fig:cavity}
\end{figure}

Even assuming that probabilities are independent, however, there is another reason why Eq.~\eqref{eq:wrong} is incorrect.  Consider Fig.~\ref{fig:cavity}, which shows a possible configuration of the network around node~$i$.  In this example, node~$i$ is connected to the giant component via one of its neighbors~$j$ and this means in turn that its other neighbor~$k$ is also connected to the giant component via~$i$.  Thus $i$ has a neighbor~$k$ in the giant component but $k$ is not, and cannot be, $i$'s~link to the giant component, since $k$ is itself only connected via~$i$.  So the question of whether $i$ is in the giant component is not merely a matter of whether $i$'s neighbor is in the giant component: $i$'s~neighbor must be connected to the giant component by some route \emph{other than via~$i$ itself.}

We can modify Eq.~\eqref{eq:wrong} to account for this by making one simple change: we remove node~$i$ from the network before applying the equation, which automatically disconnects any nodes whose only connection to the giant component was via~$i$.  Any remaining neighbors~$j$ who are still connected to the giant component must be connected by some other route and it is these neighbors that we care about.

We define a new quantity~$\mu_{i\from j}$ to be the probability that node~$j$ is not in the giant component of the network when $i$ is removed.  Then the correct equation is
\begin{equation}
\mu_i = \prod_{j\in \mathcal{N}_i} \mu_{i\from j}.
\label{eq:gc1}
\end{equation}
This equation will now correctly tell us if $i$ is in the giant component.

To use this equation we still need to compute~$\mu_{i\from j}$ itself. The probability that $j$ is not in the giant component after $i$ has been removed from the network is equal to the probability that none of $j$'s neighbors other than~$i$ is in the giant component.  Hence
\begin{equation}
\mu_{i\from j} = \prod_{\substack{k\in \mathcal{N}_j\\ k\ne i}} \mu_{j\from k}.
\label{eq:gc2}
\end{equation}
This is a \defn{message passing equation}.  We can think of the probability~$\mu_{i\from j}$ as a ``message'' that node~$i$ receives from its neighbor~$j$.  Node~$j$ transmits the probability that it is not in the giant component.  This probability is in turn calculated from the messages that $j$ receives from its other neighbors according to Eq.~\eqref{eq:gc2}.

There are two equations of the form~\eqref{eq:gc2} for every edge in the network---one going in either direction along the edge---for a total of $2m$ equations, where $m$ is the number of edges in the network.  If we can solve these~$2m$ equations for the $2m$ variables~$\mu_{i\from j}$ then we can substitute the results into Eq.~\eqref{eq:gc1} and calculate the probabilities~$\mu_i$.  In practice, the solution is normally computed by simple iteration.  We choose any suitable set of starting values for the $\mu_{i\from j}$, such as random values in $(0,1)$ for instance, and iterate~\eqref{eq:gc2} to convergence.

\begin{figure}
\begin{center}
\includegraphics[width=7.5cm]{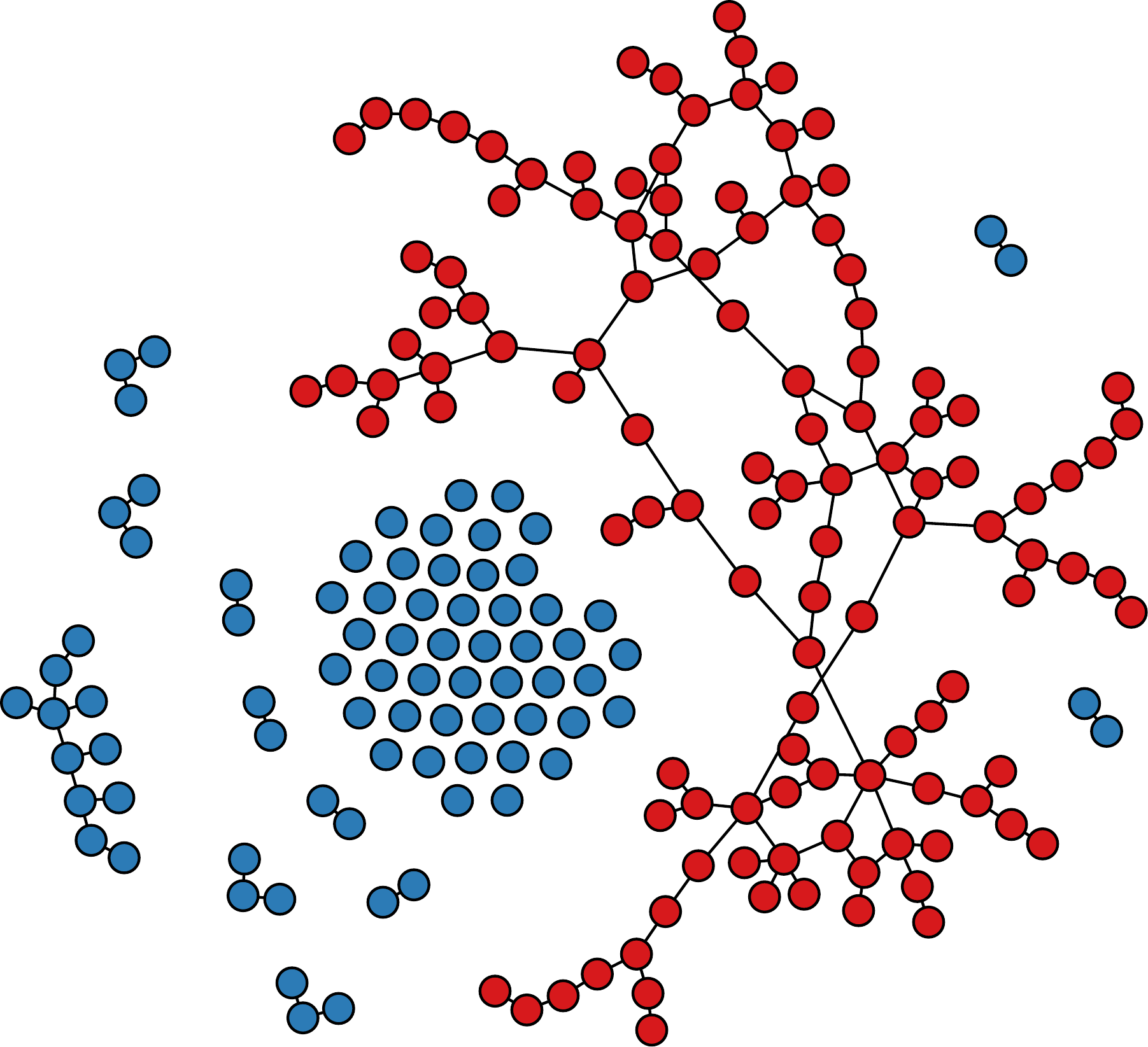}
\end{center}
\caption{Application of the method of Eqs.~\eqref{eq:gc1} and~\eqref{eq:gc2} to a small network.  The converged values of $\mu_i$ are all either zero or one and the colors of the nodes indicate the values.}
\label{fig:giantcomp}
\end{figure}

Figure~\ref{fig:giantcomp} shows what happens when we do this on a small example network.  The colors of the nodes in the figure represent the values of $\mu_i$ for each node and, as expected, there are only two values in this case, zero and one, indicating whether a node is in the giant component or not.  As we can see, the calculation has correctly identified all the nodes in the largest component of the network.

\subsection{A more substantial example: Percolation}
\label{sec:percolation}
The example of the previous section illustrates the basic message passing approach, but it is relatively trivial.  Let us see how to apply the method to a less trivial example.  \defn{Percolation} is a random process used, among other things, to model of the spread of disease over contact networks and the robustness of networks to failure of their connections~\cite{AJB00,CEBH00,Newman18c}.  In the percolation process we ``occupy'' edges in a network uniformly at random with some probability~$p$ and observe the clusters of nodes connected together by the occupied edges.  (Technically this is ``edge percolation.''  One can also study node percolation, where it is the nodes that are occupied, but we will not do so here.)  If $p$ is small there will be only small clusters, but for large enough~$p$ there will be one (and only one) giant cluster plus, potentially, some small clusters as well.  In between these two regimes lies the percolation transition, the value $p_c$ at which a giant cluster first forms.  Because of the random nature of the edge occupation process, there can be some fluctuation in~$p_c$, but the value becomes more and more narrowly concentrated as the size of the network grows.

It is only a small step from the methods of the previous section to the calculation of the probability that a node in a network belongs to the giant percolation cluster for given~$p$.  And because percolation is a probabilistic process, this is now a true probability, taking not just the values zero and one but any value in between.  Our presentation follows the line of argument given by Karrer~\etal~\cite{KNZ14} and Hamilton and Pryadko~\cite{HP14}.

Let $\mu_{i\from j}$ be the probability that node~$j$ does not belong to the giant cluster if $i$ is removed from the network.  There are two ways for $i$ to not be connected to the giant cluster via its neighbor~$j$.  The first is that the edge between $i$ and $j$ is unoccupied, which happens with probability $1-p$.  The second is that the edge is occupied (probability~$p$) but $j$ is itself not in the giant cluster (probability~$\mu_{i\from j})$.  So the total probability is $1 - p + p \mu_{i\from j}$ and the overall probability~$\mu_i$ that~$i$ is not in the giant cluster is given by a product over neighbors of~$i$ thus:
\begin{equation}
\mu_i = \prod_{j\in\mathcal{N}_i} (1 - p + p\mu_{i\from j}).
\label{eq:perc1}
\end{equation}

\begin{figure}
\begin{center}
\includegraphics[width=6.5cm]{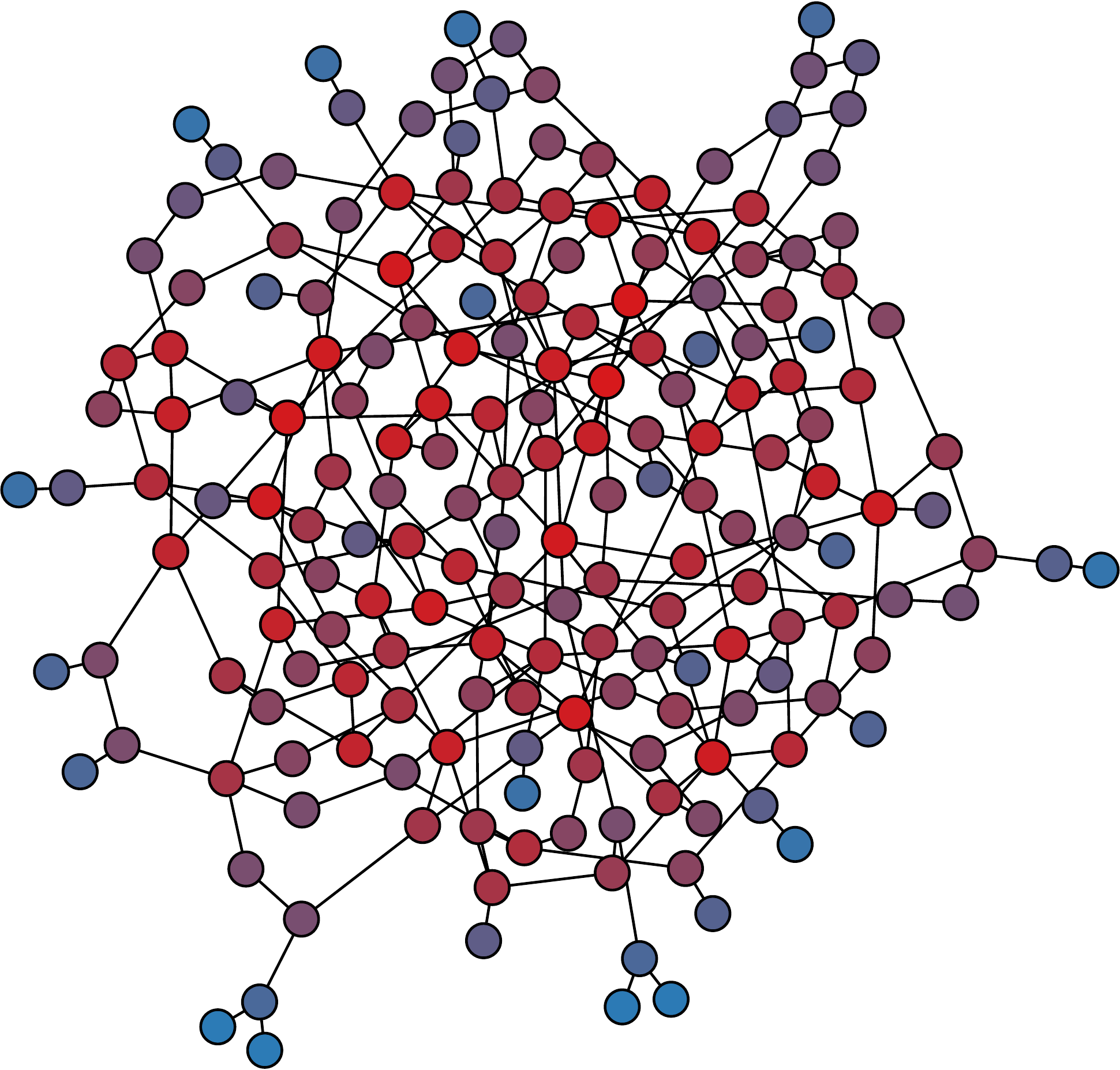}
\end{center}
\caption{Application of Eqs.~\eqref{eq:perc1} and~\eqref{eq:perc2} to a small network.  The colors of the nodes indicate the probability of belonging to the giant percolation cluster, calculated from~$\mu_i$, with red corresponding to higher probability and blue to lower.}
\label{fig:perc}
\end{figure}

Now we repeat the same argument to calculate~$\mu_{i\from j}$ itself.  The probability that $j$ is not in the giant cluster when $i$ is removed from the network is equal to the probability that $j$ is not connected to the giant cluster via any of its neighbors other than~$i$:
\begin{equation}
\mu_{i\from j} = \prod_{\substack{k\in\mathcal{N}_j\\ k\ne i}}
  (1 - p + p\mu_{j\from k}).
\label{eq:perc2}
\end{equation}
These are the message passing equations for percolation on a network and again we solve them by iteration from any suitable initial condition.  Figure~\ref{fig:perc} shows the result for a small network.  The colors of the nodes in this figure indicate the probability that a node belongs to the giant cluster.  Note how the nodes around the periphery of the network have lower probability than those in the center.

Unlike the giant component calculation of Section~\ref{sec:mp}, this percolation calculation offers something that cannot be achieved with simpler numerical means.  There exist algorithms that can find the giant cluster in a percolation system, but they do so only for one individual realization of the randomly occupied edges.  If one wanted to calculate the probability~$\mu_i$ of being in the giant cluster, one would have to run such an algorithm many times for many random realizations and average the results, which would be a complicated and computationally demanding process, and one moreover that would yield only an approximate answer, subject to statistical sampling error.  Message passing on the other hand directly yields the value of~$\mu_i$, averaged over all realizations of the randomness, in a single calculation.

\begin{figure}
\begin{center}
\includegraphics[width=\columnwidth]{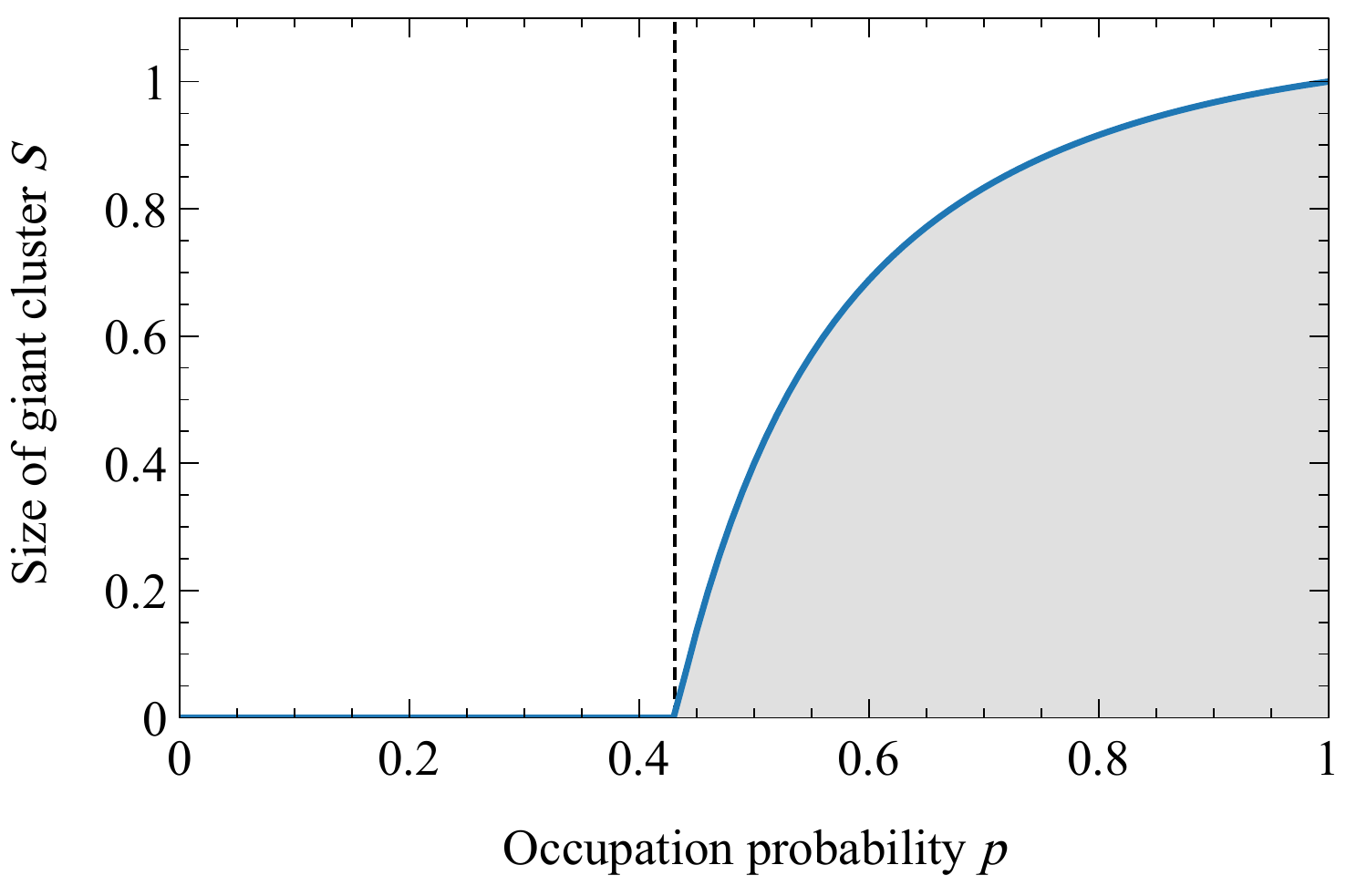}
\end{center}
\caption{The size of the giant cluster as a function of occupation probability~$p$ for edge percolation on the network of Fig.~\ref{fig:perc}, calculated using the message passing method of Eqs.~\eqref{eq:perc1}, \eqref{eq:perc2} and~\eqref{eq:S}.  The vertical dashed line indicates the expected position of the percolation transition from Eq.~\eqref{eq:pc}.}
\label{fig:percgc}
\end{figure}

One can also extend the method to calculate other properties of the percolation process, including global properties.  For instance, we can calculate the average size of the giant cluster as follows.  We define a variable~$s_i$ that is 1 if $i$ belongs to the giant cluster and 0 if it does not.  Then the probability distribution of this variable is $(1-\mu_i)_{\vphantom{i}}^{s_i} \mu_i^{1-s_i}$ and the size of giant cluster as a fraction of network size is $(1/n) \sum_i s_i$.  Hence the expected size~$S$ of the giant cluster is
\begin{align}
S &= \sum_{\set{s_i}} \biggl[ {1\over n} \sum_i s_i \biggr]
     \biggl[ \prod_i (1-\mu_i)^{s_i} \mu_i^{1-s_i} \biggr] \nonumber\\
  &= {1\over n} \sum_i \sum_{s_i=0,1} s_i (1-\mu_i)^{s_i} \mu_i^{1-s_i} \nonumber\\
  &= 1 - {1\over n} \sum_{i=1}^n \mu_i.
\label{eq:S}
\end{align}
Figure~\ref{fig:percgc} shows the value of this quantity as a function of~$p$ for the same small network as in Fig.~\ref{fig:perc}.  The figure displays the classic percolation transition behavior, with no giant cluster for small~$p$ and an abrupt phase transition---around $p=0.43$ in this case---at which a giant cluster appears.  We study this phase transition in detail in the following section.

\section{Phase transitions}
\label{sec:phase}
The message passing method is a useful tool for numerical calculation of node properties on networks as well as certain global quantities.  As mentioned in the introduction, however, message passing can also be used as the foundation for further analytic calculations, especially with regard to phase transitions in networks.  In this section we illustrate this point for the case of percolation.

As can be seen in Fig.~\ref{fig:percgc}, percolation shows a phase transition at a critical occupation probability~$p_c$ between a percolating state in which there exists a giant cluster and a non-percolating state with no giant cluster and small clusters only.  When we are below this transition, where $p<p_c$, the probability $\mu_{i\from j}$ that node~$j$ is not in the giant cluster is, by definition, 1~for all~$i,j$.  Looking at the message passing equations, Eq.~\eqref{eq:perc2}, we see that this is in fact a solution of the equations for any value of~$p$---setting $\mu_{i\from j}=1$ for all $i,j$ just gives ``$1=1$''.  So does this mean that there is never a giant cluster in the network?  It does not, because the outcome of the message passing calculation depends not only on whether this solution exists but also on whether the calculation actually returns this solution, versus some other solution.

We can think of the iteration of Eq.~\eqref{eq:perc2} as a discrete-time dynamical system that moves us around the space defined by the set of probabilities~$\mu_{i\from j}$.  The crucial question we need to answer is whether this process converges to the trivial solution $\mu_{i\from j}=1$.  If it does, then there is no giant cluster in the network.  If it converges to some other (nontrivial) solution, then there is a giant cluster.  Another way to say the same thing is that there is no giant cluster if the trivial solution is a stable fixed point of the iteration.  If it is unstable then there is a giant cluster.

This now gives us a simple method for determining whether there is a giant cluster: we linearize to find the stability of the trivial fixed point.  We write $\mu_{i\from j} = 1 - \epsilon_{i\from j}$ (with a negative sign since $\mu_{i\from j}\le1$) and substitute into Eq.~\eqref{eq:perc2} to get
\begin{align}
1 - \epsilon_{i\from j} &= \prod_{\substack{k\in\mathcal{N}_j\\ k\ne i}}
  (1 - p\epsilon_{j\from k}) \nonumber\\
  &= 1 - p \sum_{\substack{k\in\mathcal{N}_j\\ k\ne i}} \epsilon_{j\from k}
    + \Ord(\epsilon^2).
\label{eq:pc1}
\end{align}
Hence, close to the fixed point we have the linear form
\begin{equation}
\epsilon_{i\from j} = p \sum_{\substack{k\in\mathcal{N}_j\\ k\ne i}} \epsilon_{j\from k}.
\label{eq:pc2}
\end{equation}
Alternatively we can write this in matrix format as
\begin{equation}
\bm{\epsilon} = p\mat{B}\bm{\epsilon},
\label{eq:pc3}
\end{equation}
where $\bm{\epsilon}$ is the $2m$-element vector with elements~$\epsilon_{i\from j}$ and $\mat{B}$ is a $2m\times2m$ non-symmetric matrix with one row and column for each directed edge $i\from j$ and elements chosen so as to correctly reproduce Eq.~\eqref{eq:pc2}.  By inspection, this requires $B_{i\from j,k\from l} = \delta_{jk}(1-\delta_{il})$.  The resulting matrix, known as the Hashimoto edge-incidence matrix or more commonly the \defn{non-backtracking matrix}, appears in a number of contexts in network theory, including the study of community detection algorithms~\cite{Krzakala13} and centrality metrics~\cite{MZN14}.

The trivial fixed point is unstable if a small $\bm{\epsilon}$ grows under the iteration of Eq.~\eqref{eq:pc3}, which is equivalent to saying that $p\lambda>1$, where $\lambda$ is the leading eigenvalue of~$\mat{B}$.  Thus there will be a giant cluster if and only if $p>1/\lambda$ and hence we conclude that
\begin{equation}
p_c = {1\over\lambda}
\label{eq:pc}
\end{equation}
is the percolation threshold on the network~\cite{KNZ14,HP14}.  Applying this approach to the network of Fig.~\ref{fig:perc} the resulting value of~$p_c$ is shown as the vertical dashed line in Fig.~\ref{fig:percgc} and closely matches the apparent phase transition point at which the giant cluster appears.

Thus by considering the properties of the message passing equations as a dynamical system we have been able to derive a new formal result about the position of the percolation phase transition on networks.

\section{Other applications of\\message passing}
\label{sec:applications}
Message passing can be applied to a wide range of other calculations on networks.  We give three illustrative examples in this section.

\subsection{The Ising model}
\label{sec:ising}
The large class of finite-temperature spin models in physics, which includes the Ising model, the Heisenberg model, the XY model, and spin-glass and random-field models, can be treated using message passing methods~\cite{MM09,YGDM11}.  Let us take as an example the Ising model in zero magnetic field, which is defined on a general network by the Hamiltonian
\begin{equation}
H = - \tfrac12 \sum_{ij} A_{ij} s_i s_j,
\label{eq:H}
\end{equation}
where $s_i = \pm1$ is an Ising spin on node~$i$ and $A_{ij}$ is an element of the adjacency matrix~$\mat{A}$ of the network with value $A_{ij}=1$ if there exists an edge between nodes $i$ and $j$ and 0 otherwise.

We can usefully rewrite this Hamiltonian by ``centering'' it around a particular node~$i$ thus:
\begin{align}
H &= - \sum_{j\in\mathcal{N}_i} \Bigl[ s_i s_j
     + \sum_{\substack{k\in\mathcal{N}_j\\k\ne i}} \Bigl[ s_j s_k
     + \sum_{\substack{l\in\mathcal{N}_k\\l\ne j}} \Bigl[ s_k s_l
     + \ldots \Bigr] \Bigr] \Bigr] \nonumber\\
  &= - \sum_{j\in\mathcal{N}_i} \bigl[ s_i s_j + h_{i\from j}(s_j) \bigr],
\label{eq:hdecomp}
\end{align}
where
\begin{align}
h_{i\from j}(s_j) &= \sum_{\substack{k\in\mathcal{N}_j\\k\ne i}} \Bigl[ s_j s_k
     + \sum_{\substack{l\in\mathcal{N}_k\\l\ne j}} \Bigl[ s_k s_l
     + \ldots \Bigr] \Bigr] \nonumber\\
  &= \sum_{\substack{k\in\mathcal{N}_j\\k\ne i}} \bigl[ s_j s_k + h_{j\from k}(s_k)
   \bigr].
\label{eq:hij}
\end{align}
Equation~\eqref{eq:hdecomp} correctly counts each interaction $s_i s_j$ exactly once, so long as no nodes appear in more than one of the sums, which is equivalent to saying that there are no paths between the spins around the central node~$i$, other than via~$i$ itself.  As with our percolation example, we will, for the moment, assume this to be the case.  In Section~\ref{sec:loopy} we show how to remove this assumption.  Note also that while our notation $h_{i\from j}(s_j)$ explicitly includes only the dependence on~$s_j$, the value of $h_{i\from j}(s_j)$ does depend on other spins as well.  The latter however will be summed over shortly and so will play no further role.

As we have said, message passing methods are particularly useful for calculating properties of individual nodes.  Let us take as an example the calculation of the probability within the Ising model that the spin~$s_i$ has a particular value~$r=\pm1$.  At inverse temperature~$\beta$ this probability is given by the Boltzmann average
\begin{equation}
P[s_i=r] = {1\over Z} \sum_{\set{s_i}} \mathbbm{1}_{s_i=r}\,\e^{-\beta H},
\end{equation}
where $Z=\sum_{\set{s_i}} \e^{-\beta H}$ is the partition function and $\mathbbm{1}_x$ is the indicator function which is 1 if $x$ is true and 0 otherwise.  Using Eq.~\eqref{eq:hdecomp}, the sum in the numerator can be expanded as
\begin{align}
\sum_{\set{s_i}} &\mathbbm{1}_{s_i=r}\,\e^{-\beta H} = \sum_{\set{s_i}}
  \exp\biggl( \beta \sum_{j\in\mathcal{N}_i} \bigl[ rs_j + h_{i\from j}(s_j) \bigr]
  \biggr) \nonumber\\
  &= \prod_{j\in\mathcal{N}_i} \> \prod_{\substack{k\notin\mathcal{N}_i\\k\ne i}}
     \sum_{s_j} \sum_{s_k} \e^{\beta [ rs_j + h_{i\from j}(s_j) ]}
     \nonumber\\
  &= \prod_{j\in\mathcal{N}_i} \> \sum_{s_j} \e^{\beta rs_j}
     \prod_{\substack{k\notin\mathcal{N}_i\\k\ne i}} \sum_{s_k}
      \e^{\beta h_{i\from j}(s_j)}.
\end{align}
Now we define a message~$\mu^s_{i\from j}$ according to
\begin{equation}
\mu^s_{i\from j} = {1\over Z_{i\from j}}
  \prod_{\substack{k\notin\mathcal{N}_i\\k\ne i}} \sum_{s_k} \e^{\beta h_{i\from j}(s)},
\label{eq:mij}
\end{equation}
where $Z_{i\from j}$ is a normalizing factor chosen to make $\sum_{s=\pm1} \mu^s_{i\from j} = 1$.  Then
\begin{align}
P[s_i=r] &= {1\over Z} \prod_{j\in\mathcal{N}_i} \> \sum_s \e^{\beta rs}
           Z_{i\from j} \mu^s_{i\from j} \nonumber\\
  &= {1\over Z_i} \prod_{j\in\mathcal{N}_i}
     \bigl( \e^{\beta r} \mu^+_{i\from j} + \e^{-\beta r} \mu^-_{i\from j} \bigr),
\label{eq:marginal}
\end{align}
with $\mu^\pm_{i\from j}$ denoting the value of $\mu^s_{i\from j}$ for $s=\pm1$ and
\begin{equation}
Z_i = {Z\over\prod_{j\in\mathcal{N}_i} Z_{i\from j}}.
\label{eq:zzz}
\end{equation}
The partition function~$Z$ is difficult to calculate directly, but evaluating~\eqref{eq:marginal} only requires~$Z_i$, which can be done easily by noting that $\sum_{r=\pm1} P[s_i=r] = 1$, so
\begin{equation}
Z_i = \sum_{r=\pm1} \prod_{j\in\mathcal{N}_i}
     \bigl( \e^{\beta r} \mu^+_{i\from j} + \e^{-\beta r} \mu^-_{i\from j} \bigr).
\label{eq:zi}
\end{equation}

To use Eq.~\eqref{eq:marginal} we still need the values of the messages~$\mu^s_{i\from j}$, which we can calculate from Eqs.~\eqref{eq:hij} and~\eqref{eq:mij} thus:
\begin{align}
\mu^r_{i\from j} &= {1\over Z_{i\from j}}
    \prod_{\substack{k\notin\mathcal{N}_i\\k\ne i}} \sum_{s_k}
     \e^{\beta h_{i\from j}(r)} \nonumber\\
  &= {1\over Z_{i\from j}} \prod_{\substack{k\in\mathcal{N}_j\\k\ne i}}
     \prod_{\substack{l\notin\mathcal{N}_j\\l\ne j}}
     \sum_{s_k} \sum_{s_l}
     \e^{\beta [r s_k + h_{j\from k}(s_k)]} \nonumber\\
  &= {1\over Z_{i\from j}} \prod_{\substack{k\in\mathcal{N}_j\\k\ne i}} \sum_{s_k}
     \e^{\beta r s_k}
     \prod_{\substack{l\notin\mathcal{N}_j\\l\ne j}} \sum_{s_l}
     \e^{\beta h_{j\from k}(s_k)} \nonumber\\
  &= {1\over Z_{i\from j}} \prod_{\substack{k\in\mathcal{N}_j\\k\ne i}} \sum_s
     \e^{\beta rs} \mu^s_{j\from k},
\end{align}
or
\begin{equation}
\mu^r_{i\from j} = {1\over Z_{i\from j}} \prod_{\substack{k\in\mathcal{N}_j\\k\ne i}}
     \bigl( \e^{\beta r} \mu^+_{j\from k} + \e^{-\beta r} \mu^-_{j\from k} \bigr).
\label{eq:isingmp1}
\end{equation}
Since $Z_{i\from j}$ is defined to make $\sum_{r=\pm1} \mu^r_{i\from j} = 1$, its value is given by
\begin{equation}
Z_{i\from j} = \sum_{r=\pm1}\>\prod_{\substack{k\in\mathcal{N}_j\\k\ne i}}
  \bigl( \e^{\beta r} \mu^+_{j\from k} + \e^{-\beta r} \mu^-_{j\from k} \bigr).
\label{eq:zij1}
\end{equation}

Equations~\eqref{eq:isingmp1} and~\eqref{eq:zij1} define the message passing process for the Ising model and as before can be solved by simple iteration, starting for instance from random values.  Once the values converge, we can calculate the probability that~$s_i = \pm1$ from Eq.~\eqref{eq:marginal}, and from these probabilities we can calculate other quantities of interest.  For example, the average magnetization $m_i = \av{s_i}$ at node~$i$ is given by $m_i = P[s_i=+1] - P[s_i=-1]$ and from this one can calculate the global average magnetization $m = (1/n) \sum_i m_i$.

The message passing equations also give us a way to calculate the partition function~$Z$ of the Ising model, something that is difficult to do by other methods such as Monte Carlo simulation.  Rearranging Eq.~\eqref{eq:zzz} we have
\begin{equation}
Z = Z_i\prod_{j\in\mathcal{N}_i} Z_{i\from j},
\end{equation}
so we can calculate the partition function directly from the quantities $Z_i$ and~$Z_{i\from j}$, which are evaluated as part of the message passing process.  From $Z$ we can then calculate other quantities such as the free energy, which is $F = -\beta^{-1} \log Z = -\beta^{-1} (\log Z_i + \sum_{j\in\mathcal{N}_i} \log Z_{i\from j})$ or, if we prefer a more symmetric formulation,
\begin{equation}
F = -{1\over n\beta} \biggl( \sum_i \log Z_i + \sum_{i\from j} \log Z_{i\from j} \biggr),
\end{equation}
where the second sum is over all directed edges $i\from j$ in the network.

\begin{figure}
\begin{center}
\includegraphics[width=\columnwidth]{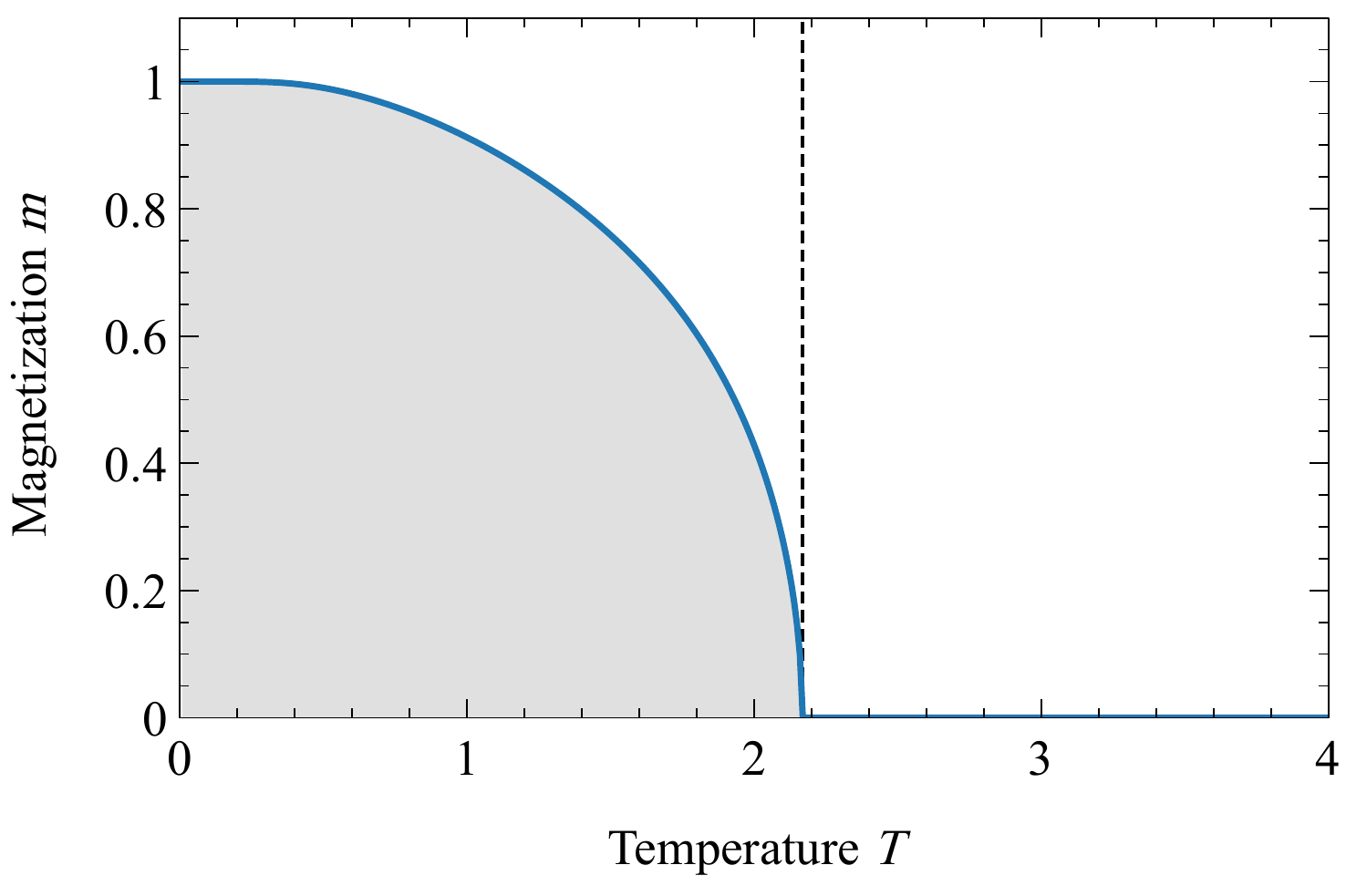}
\end{center}
\caption{The magnetization of an Ising model on a small network as a function of temperature~$T=1/\beta$, calculated from Eqs.~\eqref{eq:marginal} and~\eqref{eq:isingmp1}.  The vertical dashed line indicates the expected position of the ferromagnetic phase transition, calculated from the leading eigenvalue of the non-backtracking matrix using Eq.~\eqref{eq:isingtc}.}
\label{fig:ising}
\end{figure}

Figure~\ref{fig:ising} shows results for average magnetization calculated by message passing on a small network.  The Ising model in zero field is up-down symmetric, so one might imagine the value of~$m$ would just be zero, but this is not the case.  The message passing calculation displays spontaneous symmetry breaking, just as the Ising model itself does.  Zero magnetization implies that all spins are equally likely to be up or down, so $\mu^+_{i\from j}=\mu^-_{i\from j}=\frac12$ for all $i,j$, and Eq.~\eqref{eq:isingmp1} does have such a solution, but it also has two other solutions, one where we have $\mu^+_{i\from j}>\frac12$ and $\mu^-_{i\from j}<\frac12$ on average and one where we have the reverse, and for these solutions $m$ is nonzero.  As a simple example, in the low-temperature limit $\beta\to\infty$ it is straightforward to see that $\mu^\pm_{i\from j} = 1$ and 0 are solutions, which correspond to all spins up and all spins down.

Which solution the message passing process converges to depends on whether the solutions are stable or unstable under the iteration of Eq.~\eqref{eq:isingmp1}.  In particular, if the symmetric solution at $\mu^+_{i\from j} = \mu^-_{i\from j} = \frac12$ is unstable then we must converge to one of the other, symmetry-broken solutions and hence we get a nonzero magnetization, indicating that we are below the symmetry-breaking phase transition, in the ferromagnetic regime of the model.  Conversely, if we converge to the symmetric solution then we are above the transition.  As with percolation, therefore, the point at which the symmetric solution for the Ising model undergoes a bifurcation from stable to unstable corresponds to the phase transition.

We can check for stability by once again expanding about the symmetric point.  Observing that $\mu^-_{i\from j} = 1 - \mu^+_{i\from j}$ and putting $\mu^\pm_{i\from j} = \frac12(1\pm\epsilon_{i\from j})$ in Eq.~\eqref{eq:isingmp1} we have
\begin{align}
\tfrac12(1+&\,\epsilon_{i\from j}) = {1\over Z_{i\from j}}
     \prod_{\substack{k\in\mathcal{N}_j\\k\ne i}} \!
     \tfrac12 \bigl[ \e^\beta (1\!+\epsilon_{j\from k})
                     + \e^{-\beta} (1\!-\epsilon_{j\from k}) \bigr] \nonumber\\
  &= {1\over Z_{i\from j}} \prod_{\substack{k\in\mathcal{N}_j\\k\ne i}}
     \bigl( \cosh\beta + \epsilon_{j\from k} \sinh\beta \bigr) \nonumber\\
  &= {(\cosh\beta)^{|\mathcal{N}_j|-1}\over Z_{i\from j}}
     \biggl[ 1 + \tanh\beta \sum_{\substack{k\in\mathcal{N}_j\\k\ne i}} \epsilon_{j\from k}
     \biggr] + \Ord(\epsilon^2),
\end{align}
where $|\mathcal{N}_j|$ is the degree of node~$j$, i.e.,~the number of its neighbors.  Making the same substitution in Eq.~\eqref{eq:zij1} gives just $Z_{i\from j} = 2 (\cosh\beta)^{|\mathcal{N}_j|-1}$, and hence, neglecting terms of order~$\epsilon^2$, the linearized message passing equations are
\begin{equation}
\epsilon_{i\from j} = \tanh\beta \sum_{\substack{k\in\mathcal{N}_j\\k\ne i}} \epsilon_{j\from k}.
\end{equation}
This has the same form as Eq.~\eqref{eq:pc2} for the percolation case and carries the same implication: just as the critical occupation probability for percolation satisfies $\lambda p_c=1$, where $\lambda$ is the leading eigenvalue of the non-backtracking matrix, so the critical temperature of the Ising model satisfies $\lambda\tanh\beta_c=1$, or
\begin{equation}
\beta_c = \mathop{\textrm{arctanh}} {1\over\lambda}.
\label{eq:isingtc}
\end{equation}
The corresponding value of the critical temperature $T_c = 1/\beta_c$ is marked as the vertical dashed line in Fig.~\ref{fig:ising} and agrees nicely with the apparent transition point between $m>0$ and $m=0$.

\subsection{Graph spectra}
\label{sec:spectra}
The eigenvalue spectrum of the adjacency matrix of a graph or network plays a central role in several aspects of the theory of networks, including the calculation of centrality measures~\cite{Bonacich87}, behavior of dynamical systems on networks~\cite{PG16}, percolation properties~\cite{KNZ14}, community detection~\cite{Fortunato10}, and network epidemiology~\cite{Newman18c}.  Complete matrix spectra can be calculated numerically by standard means such as the QR algorithm~\cite{PTVF92}, but this approach is slow and limited to relatively small networks, up to about 10\,000 nodes.  As another example of message passing, we here show how one can use the method to calculate graph spectra in very competitive running times, especially on sparse graphs.  Our presentation follows that of~\cite{RCKT08,NZN19}.

One can write the spectrum of a network in terms of the spectral density
\begin{equation}
\rho(x) = {1\over n} \sum_{i=1}^n \delta(x-\lambda_i),
\label{eq:density}
\end{equation}
where $\lambda_i$ are the eigenvalues of the adjacency matrix and $\delta(x)$ is the Dirac delta function.  There are a variety of analytic forms for the delta function, but for our purposes a useful one is the Lorentzian or Cauchy distribution in the limit of zero width:
\begin{equation}
\delta(x) = \lim_{\eta\to0^+} {\eta/\pi\over x^2+\eta^2}
  = -{1\over\pi} \lim_{\eta\to0^+} \Im {1\over x+\ii\eta},
\label{eq:delta}
\end{equation}
where $\eta\to0^+$ means that $\eta$ tends to zero from above.  Substituting for $\delta(x)$ in Eq.~\eqref{eq:density} we then have
\begin{equation}
\rho(x) = -{1\over n\pi} \lim_{\eta\to0^+} \Im \sum_{i=1}^n
           {1\over x-\lambda_i+\ii\eta}.
\label{eq:realrho}
\end{equation}
In our calculations we will focus on the complex generalization of the spectral density
\begin{equation}
\rho(z) = -{1\over n\pi} \sum_{i=1}^n {1\over z-\lambda_i}
        = -{1\over n\pi} \Tr (z\mat{I}-\mat{A})^{-1},
\label{eq:complexrho}
\end{equation}
where $\mat{I}$ is the $n\times n$ identity matrix and $\mat{A}$ is the adjacency matrix as previously.  If we can calculate this quantity then the real spectral density~$\rho(x)$ of Eq.~\eqref{eq:realrho} is given by setting $z = x + \ii\eta$, taking the imaginary part, and then taking the limit as $\eta\to0^+$.

Expanding the matrix inverse in~\eqref{eq:complexrho} 
as a power series in~$z$ and taking the trace term by term gives
\begin{equation}
\rho(z) = -{1\over n\pi z} \sum_{r=0}^\infty {\Tr \mat{A}^r\over z^r}.
\label{eq:rhotrace}
\end{equation}
But the trace of $\mat{A}^r$ is equal to the number of closed (potentially self-intersecting) walks of length~$r$ on a network---walks that start and end at the same node and traverse exactly~$r$ edges along the way.  So if we can count closed walks we can evaluate Eq.~\eqref{eq:rhotrace}, and hence calculate the spectral density of our network.

To count closed walks, we first consider the more limited problem of counting closed walks that start from a specified node and return to the same node for the first and only time on their final step.  Walks that return only once like this we call \defn{excursions}.  Let us denote by $n^{(r)}_{i\from j}$ the number of excursions that start from node~$i$, take their first step along the edge from $i$ to~$j$, and return to $i$ after exactly $r$ steps.  If the neighbors of $i$ are connected to one another other than via~$i$ itself, either directly by edges or by other short paths, then it is possible for an excursion to return to $i$ along a different edge from the one it left by.  As with our other message passing calculations we will assume for the moment that this is not the case.  (Again, we will show in Section~\ref{sec:loopy} how to relax this assumption.)  Thus, all excursions that start by traversing the edge $(i,j)$ return along this same edge.

\begin{figure}
\begin{center}
\includegraphics[width=5.2cm]{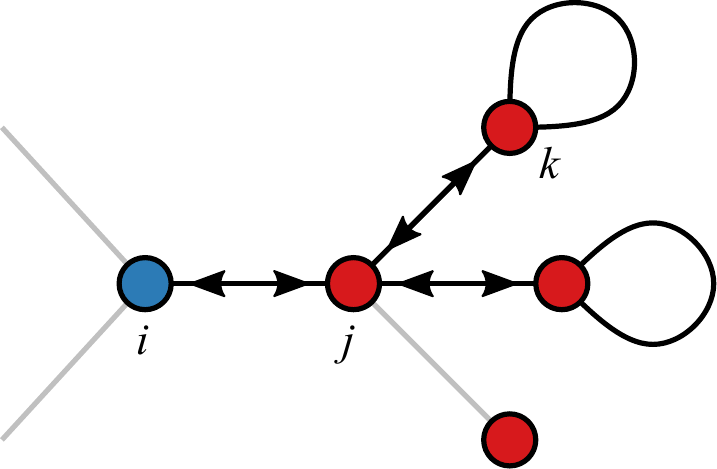}
\end{center}
\caption{An excursion from node~$i$ traverses the edge from $i$ to one of its neighbors~$j$, then makes any number of further excursions via other neighbors~$k$ of node~$j$ before traversing the edge from $j$ to $i$ again and ending.}
\label{fig:excursion}
\end{figure}

Generically, every excursion of length~$r$ now has the structure shown in Fig.~\ref{fig:excursion}: it first traverses the edge from $i$ to~$j$, then it makes some number~$m$ (possibly zero) of separate excursions from~$j$ via neighbors~$k$ (not including~$i$) and then it traverses the edge from $j$ to~$i$ back again and stops at $i$ after a total of $r$ steps.  This observation allows us to express the number of excursions $n^{(r)}_{i\from j}$ self-consistently in the form
\begin{align}
n^{(r)}_{i\from j} &= \sum_{m=0}^\infty \Biggl[ \sum_{\substack{k_1\in\mathcal{N}_j\\k_1\ne i}} \ldots
  \sum_{\substack{k_m\in\mathcal{N}_j\\k_m\ne i}} \Biggr] \Biggl[ \sum_{r_1=1}^\infty \ldots
  \sum_{r_m=1}^\infty \Biggr] \nonumber\\
  &\hspace{4em}{}\times
   \mathbbm{1}_{\sum_u r_u=r-2} \prod_{u=1}^m n^{(r_u)}_{j\from k_u}.
\end{align}
In this expression the product $\prod_{u=1}^m n^{(r_u)}_{j\from k_u}$ represents the number of combinations of excursions from~$j$ that have lengths~$r_1\ldots r_m$ and proceed via neighbors~$k_1\ldots k_m$ of~$j$ (not necessarily distinct).  The lengths and first steps are summed over all possibilities and the indicator function ensures that only those combinations whose lengths add up to $r-2$ are allowed, the remaining two steps being reserved for traversing the edge $(i,j)$ in either direction.

Now we define a message $\mu_{i\from j}(z)$ by
\begin{align}
\mu_{i\from j}&(z) = \sum_{r=1}^\infty {n^{(r)}_{i\from j}\over z^r} \nonumber\\
  &= \sum_{r=1}^\infty {1\over z^r} \sum_{m=0}^\infty
  \Biggl[ \sum_{\substack{k_1\in\mathcal{N}_j\\k_1\ne i}} \ldots
  \sum_{\substack{k_m\in\mathcal{N}_j\\k_m\ne i}} \Biggr]
  \Biggl[ \sum_{r_1=1}^\infty \ldots \sum_{r_m=1}^\infty \Biggr] \nonumber\\
  &\hspace{8em}{}\times
   \mathbbm{1}_{\sum_u r_u=r-2} \prod_{u=1}^m n^{(r_u)}_{j\from k_u} \nonumber\\
  &= {1\over z^2} \sum_{m=0}^\infty
  \Biggl[ \sum_{\substack{k_1\in\mathcal{N}_j\\k_1\ne i}} \!\!\ldots\!\!\! \sum_{\substack{k_m\in\mathcal{N}_j\\k_m\ne i}} \Biggr]
  \Biggl[ \sum_{r_1=1}^\infty \!\ldots\! \sum_{r_m=1}^\infty \Biggr]
   \prod_{u=1}^m {n^{(r_u)}_{j\from k_u}\over z^{r_u}} \nonumber\\
  &= {1\over z^2} \sum_{m=0}^\infty \> \prod_{u=1}^m \Biggl[
  \sum_{\substack{k_u\in\mathcal{N}_j\\k_u\ne i}}\>\sum_{r_u=1}^\infty {n^{(r_u)}_{j\from k_u}\over z^{r_u}}
  \Biggr] \nonumber\\
  &= {1\over z^2} \sum_{m=0}^\infty \Biggl[
  \sum_{k\in\mathcal{N}_j, k\ne i} \!\!\mu_{j\from k}(z) \Biggr]^m
  \nonumber\\
  &= {1/z^2\over 1-\sum_{k\in\mathcal{N}_j, k\ne i} \mu_{j\from k}(z)}.
\label{eq:specmp1}
\end{align}
This is our message passing equation, which we solve in the normal fashion by iteration.  Once we have the values of the messages we can write the total number $n^{(r)}_i$ of closed walks of length~$r$ starting and ending at node~$i$ as a composition of any number $m$ of excursions via the nodes in $i$'s neighborhood:
\begin{align}
n^{(r)}_i &= \sum_{m=0}^\infty \biggl[ \sum_{j_1\in\mathcal{N}_i} \ldots
  \sum_{j_m\in\mathcal{N}_i} \biggr] \biggl[ \sum_{r_1=1}^\infty \ldots
  \sum_{r_m=1}^\infty \biggr] \nonumber\\
  &\hspace{4em}{}\times
   \mathbbm{1}_{\sum_u r_u=r} \prod_{u=1}^m n^{(r_u)}_{j\from k_u}.
\end{align}
Then, by a derivation similar to that of~\eqref{eq:specmp1}, we have
\begin{align}
\sum_{r=0}^\infty {\Tr \mat{A}^r\over z^r}
  &= n + \sum_{r=1}^\infty\,\sum_{i=1}^n {n^{(r)}_i\over z^r} \nonumber\\
  &= n + \sum_{i=1}^n {1\over1-\sum_{j\in\mathcal{N}_i} \mu_{i\from j}(z)}.
\end{align}
And substituting this result into Eq.~\eqref{eq:rhotrace} and neglecting the initial $+n$, which will disappear anyway when we take the imaginary part, we get our expression for the complex spectral density
\begin{equation}
\rho(z) = - {1\over n\pi z} \sum_{i=1}^n
     {1\over1-\sum_{j\in\mathcal{N}_i} \mu_{i\from j}(z)}.
\label{eq:specmp2}
\end{equation}
Taking the imaginary part now gives us~$\rho(x)$.

\begin{figure}
\begin{center}
\includegraphics[width=\columnwidth]{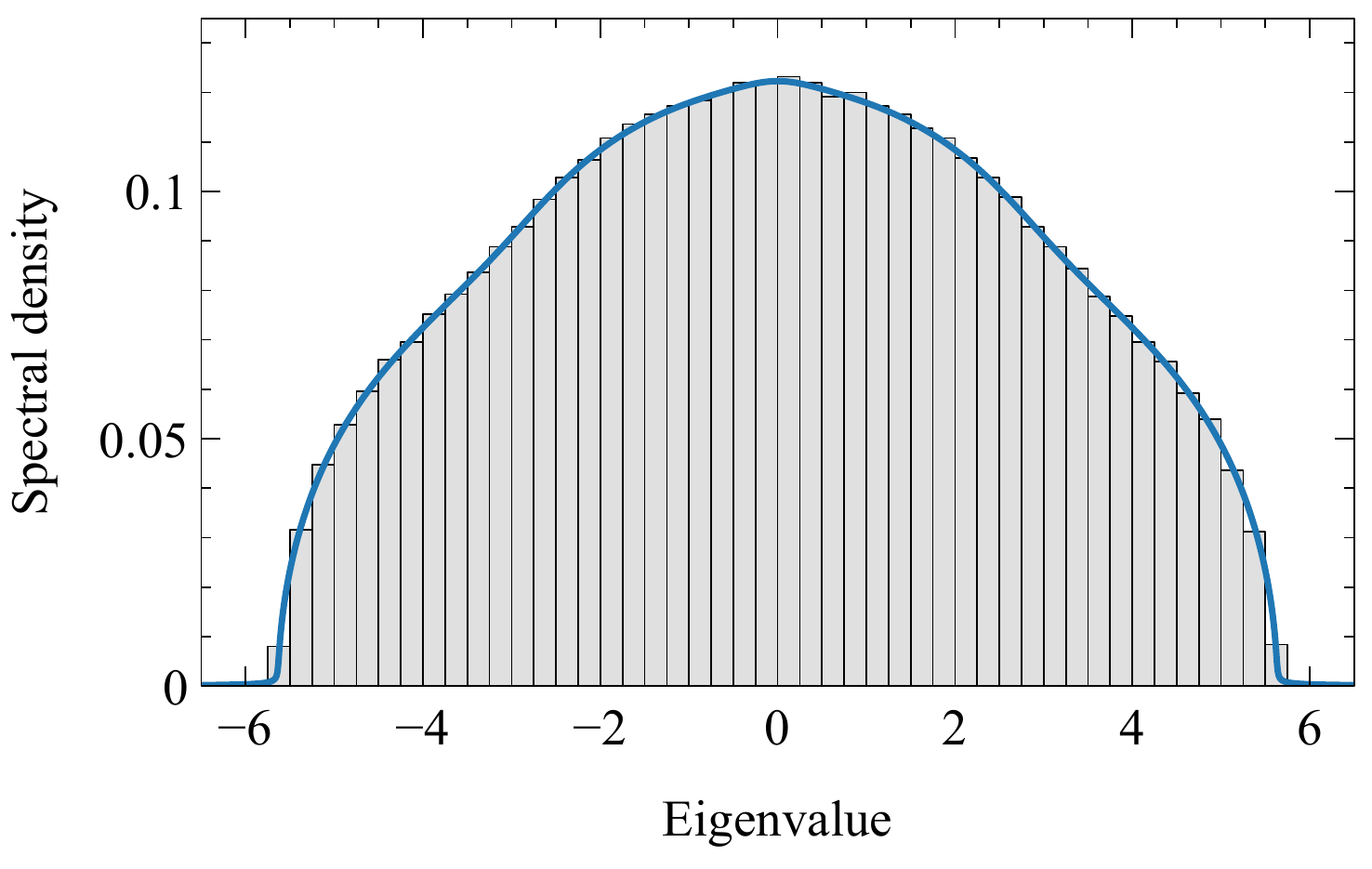}
\end{center}
\caption{The spectral density of a $10\,000$-node network calculated in two ways: first using the standard QR algorithm to compute all eigenvalues and then calculating a histogram of the results (gray), and second using the message passing equations, \eqref{eq:specmp1}~and~\eqref{eq:specmp2}, with $\eta = 0.01$ (solid curve).}
\label{fig:spect}
\end{figure}

This procedure gives us a complete recipe for calculating the spectral density of a network using message passing.  We can use it for numerical calculations by setting $z = x + \ii\eta$ for small but nonzero~$\eta$, explicitly iterating to convergence the message passing equations of~\eqref{eq:specmp1} with this value of~$z$ and taking the imaginary part of~\eqref{eq:specmp2} to find the spectral density at~$x$.  Figure~\ref{fig:spect} shows an example application to a network of $n=10\,000$ nodes.  Also shown in the figure are the results of direct numerical calculations using the QR algorithm and the agreement between the two is good.

As with our other message passing methods, one can also use the equations for the spectral density as the basis for further analytic computations.  To give a simple example, consider a random $d$-regular network, i.e.,~a network in which every node has $d$ edges but the nodes are otherwise connected together at random.  When the number of nodes~$n$ is large, the neighborhood of every node in such a network looks identical: each node has $d$ neighbors, each of which has $d$ neighbors, and so on.  This means that the messages $\mu_{i\from j}$ on all edges have the same value~$\mu(z)$ for any~$z$, so Eq.~\eqref{eq:specmp2} becomes
\begin{equation}
\rho(z) = - {1\over\pi z[1-d\mu(z)]}\,,
\label{eq:specmp3}
\end{equation}
and the message passing equations~\eqref{eq:specmp1} take the simple form
\begin{equation}
\mu(z) = {1/z^2\over1-(d-1)\mu(z)}.
\end{equation}
This last result can be rearranged into the quadratic equation $(d-1) \mu^2 - \mu + 1/z^2 = 0$ and, computing the solutions, substituting into Eq.~\eqref{eq:specmp3}, and taking the imaginary part, we find that
\begin{equation}
\rho(x) = (d/2\pi) {\sqrt{4(d-1)-x^2}\over d^2-x^2}.
\end{equation}
This is the well-known spectrum of Kesten and McKay for the random regular graph~\cite{McKay81}, but derived here by a different---and shorter---method than the traditional one.

The same approach can be extended to other model networks as well.  For instance, it can be easily modified for networks in which the number of edges at nodes---their degree---alternates between two different values, or more generally to the large class of ``equitable random graphs,'' which can have nodes of many different degrees~\cite{NM14}.  By making some further approximations, the same method can also be used to calculate the spectrum of the configuration model~\cite{NZN19}, perhaps the most important structural model in the theory of networks~\cite{MR95,NSW01}.

\subsection{Community detection}
\label{sec:comm}
\begin{figure}
\begin{center}
\includegraphics[width=8cm]{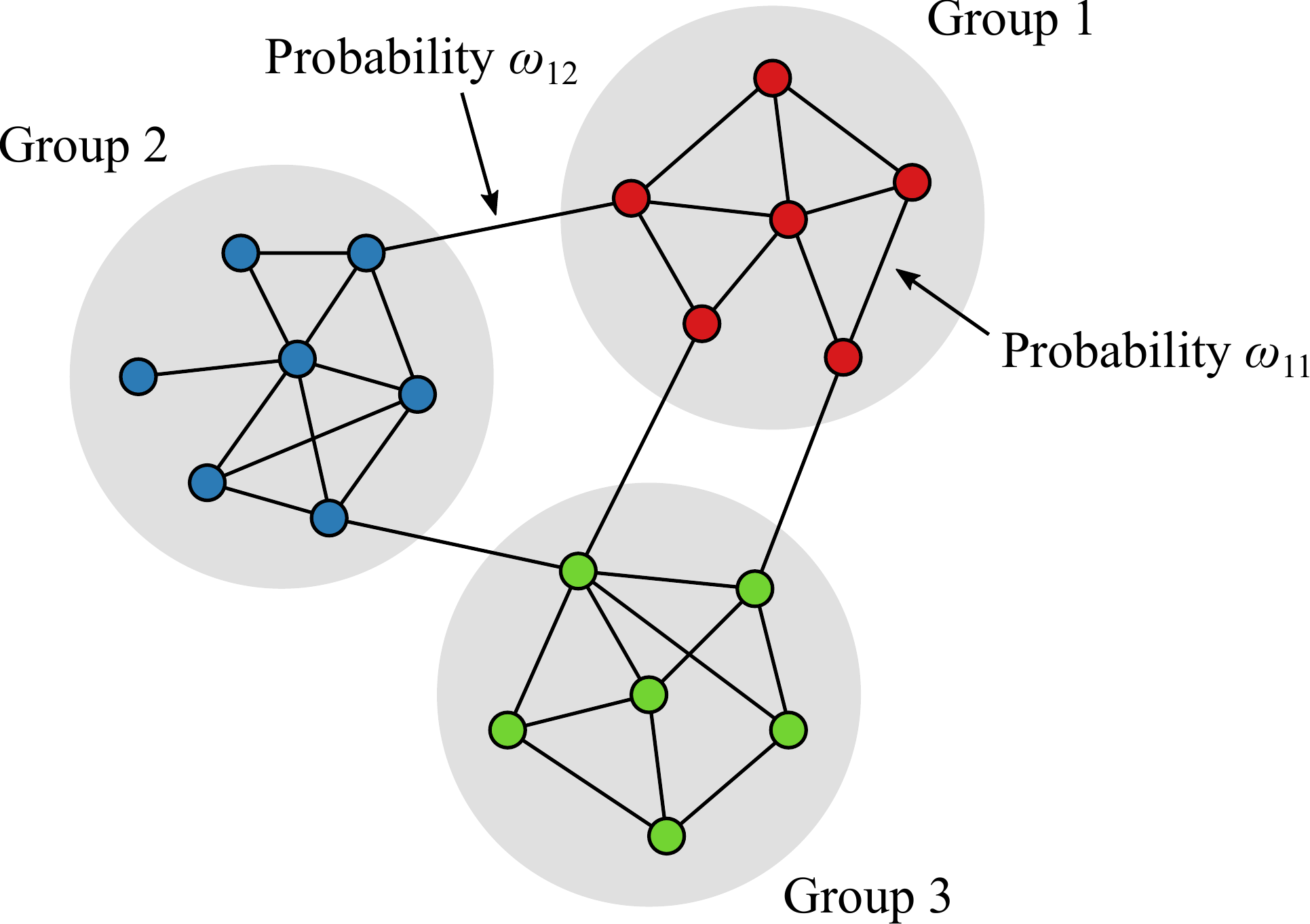}
\end{center}
\caption{Many networks display community structure, in which the network divides into tightly-knit groups of nodes, with many connections within groups but few connections between them.  The stochastic block model imitates this structure by placing edges randomly between node pairs with probabilities~$\omega_{rs}$ that depend on the groups~$r,s$ that the nodes belong to.}
\label{fig:community}
\end{figure}

As our final example of message passing we look at an application to one of the classic problems in the study of networks: community detection.  Many real-world networks have large-scale structure of the kind sketched in Fig.~\ref{fig:community}, in which the network nodes are divided into some number of groups or communities, with dense connections within groups but only sparser connections between groups.  Structure of this kind often reflects functional divisions between nodes, and the ability to find such structure in unlabeled network data can be an invaluable tool for understanding the link between form and function in networked systems~\cite{Fortunato10}.

There are a large number of techniques for detecting communities in networks, but one of the most promising, and also most mathematically principled, makes use of a fit to a statistical model of community structure, as illustrated in Fig.~\ref{fig:community}.  Briefly, suppose we take $n$ nodes and divide them among~$q$ groups numbered $1\ldots q$.  Then we place edges at random such that each pair of nodes is connected independently with probability~$\omega_{rs}$ which depends on the groups $r$ and $s$ that the nodes belong to.  Thus two nodes in group~1 would be connected with probability~$\omega_{11}$ and nodes in groups 1 and 2 would be connected with probability~$\omega_{12}$.  This model of a network is known as the \defn{stochastic block model}~\cite{HLL83}, and if the diagonal probabilities~$\omega_{rr}$ are larger than the off-diagonal ones it generates a random network with classic community structure of the kind shown in the figure.

The community detection method that we focus on here involves fitting this model (or one of its variants~\cite{ABFX08,KN11a,Peixoto17}) to observed network data by the method of maximum a posteriori probability.  Suppose we observe a network with adjacency matrix~$\mat{A} = [A_{ij}]$, which we hypothesize was generated from the stochastic block model.  If $s_i$ denotes the community to which node~$i$ belongs then the probability of generating this network is
\begin{equation}
P[\mat{A}|\vec{s}] = \prod_{i<j} \omega_{s_is_j}^{A_{ij}}
   (1-\omega_{s_is_j})^{1-A_{ij}},
\end{equation}
where $\vec{s} = [s_i]$ is the vector of community assignments and the sum over values $i<j$ ensures that we count each node pair only once.  By Bayes' rule we now have
\begin{align}
P[\vec{s}|\mat{A}] &= {P[\mat{A}|\vec{s}] P[\vec{s}]\over P[\mat{A}]}
  \nonumber\\
  &= {1\over Z} \prod_i \pi_{s_i}
     \prod_{i<j} \omega_{s_is_j}^{A_{ij}} (1-\omega_{s_is_j})^{1-A_{ij}},
\label{eq:sbmpost1}
\end{align}
where for convenience we have defined $Z = P[\mat{A}]$ and we assume a categorical prior~$P[\vec{s}]$ on the group assignments in which each node is assigned to group~$s$ with prior probability~$\pi_s$ which we choose.

Fitting the model to the observed network involves finding the most likely values of both the community assignments~$s_i$ and the parameters~$\pi_r,\omega_{rs}$ by maximizing~\eqref{eq:sbmpost1}.  Finding the parameter values is a relatively straightforward task---it can be accomplished using a standard expectation-maximization (EM) algorithm.  Finding the community assignments, on the other hand, is more difficult and it is this task that we address here, assuming that we already know the values of the parameters.

We focus on calculating the probability $q_i^s = P[s_i=r]$ that node~$i$ is assigned to group~$r$.  We will not go into the derivation of the message passing equations in detail---see Refs.~\cite{DKMZ11a,Moore17} for a discussion---but, briefly, we define a set of messages~$\mu_{i\from j}^r$ that satisfy the message passing equations
\begin{equation}
\mu_{i\from j}^r = {\pi_r\over Z_{i\from j}}
  \prod_k \Bigl( 1 - \sum_s q_k^s \omega_{rs} \Bigr)
  \prod_{\substack{k\in\mathcal{N}_j\\k\ne i}} \sum_s \omega_{rs} \mu_{j\from k}^s.
\label{eq:commmp2}
\end{equation}
In terms of these messages
\begin{equation}
q_i^r = {\pi_r\over Z_i} \prod_j \Bigl( 1 - \sum_s q_j^s \omega_{rs} \Bigr)
    \prod_{j\in\mathcal{N}_i} \sum_s \omega_{rs} \mu^s_{i\from j},
\label{eq:commmp1}
\end{equation}
while the normalizing factors $Z_i$ and $Z_{i\from j}$ are chosen so that $\sum_r q_i^r = 1$ and $\sum_r \mu_{i\from j}^r = 1$ for all $i,j$.

\begin{figure}
\begin{center}
\includegraphics[width=\columnwidth]{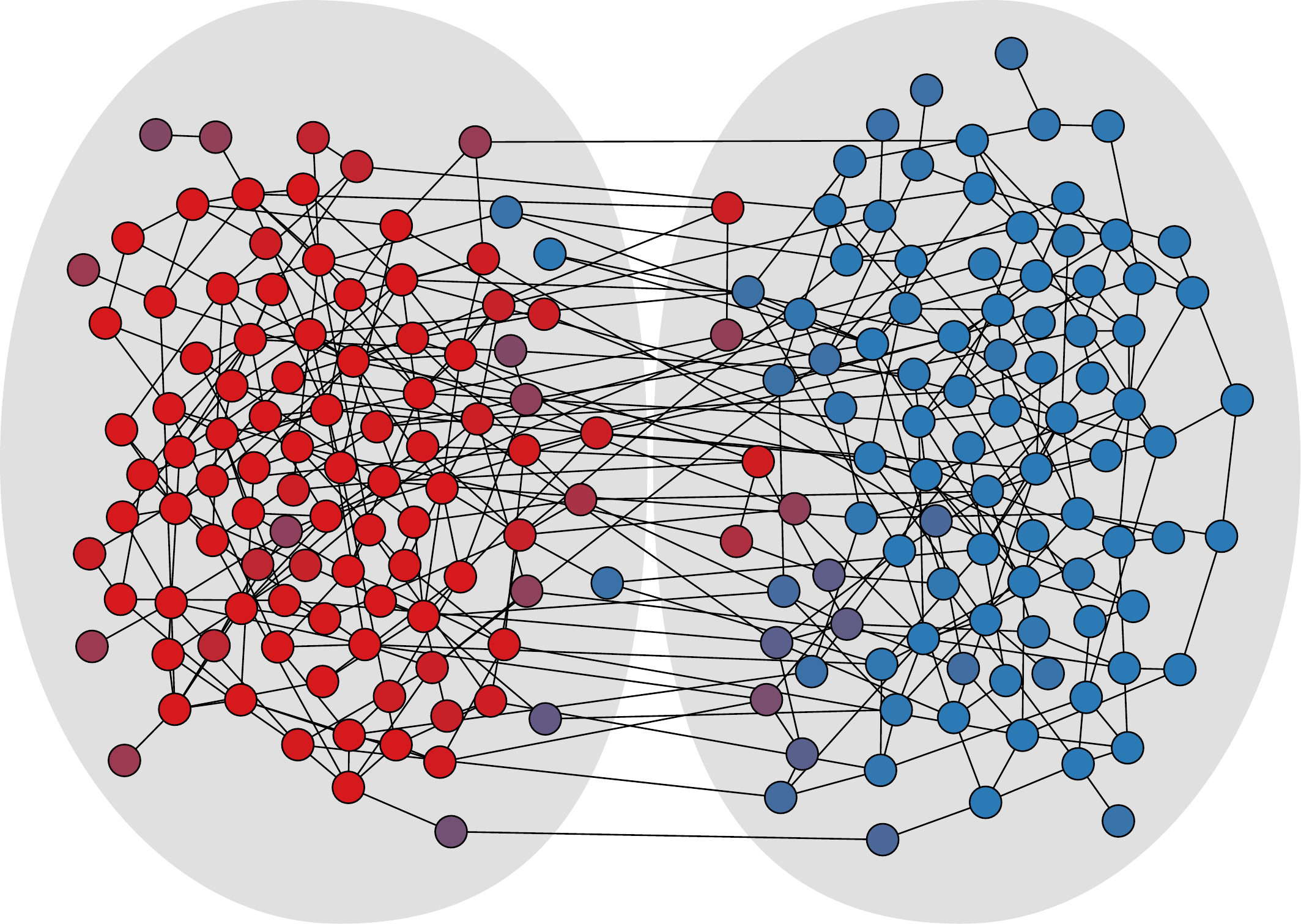}
\end{center}
\caption{A small network with two communities, indicated by the shaded regions.  The colors of the nodes correspond to the values of the message passing probabilities~$q_i^r$ from Eq.~\eqref{eq:commmp1} and in this case the message passing algorithm has successfully found the two communities, with only a few nodes incorrectly assigned.}
\label{fig:commres}
\end{figure}

Equation~\eqref{eq:commmp2} is solved by simple iteration, starting for instance from random values, then the results are substituted into Eq.~\eqref{eq:commmp1} to calculate~$q_i^r$.  Figure~\ref{fig:commres} shows an example application to a small network with two communities.  The colors of the nodes indicate the probability that the node belongs to group~1 (red) or group~2 (blue).  As we can see, the method has ably found the two groups of nodes in the network, with only a few nodes incorrectly assigned.

Message passing provides an efficient and practical algorithm for community detection in networks, but perhaps the most interesting result to come out of this approach is a more formal one.  Like the other message passing applications we have seen, the equations for community detection can be used as a starting point for further calculations, particularly focusing on phase transition behavior.  Decelle~\etal~\cite{DKMZ11a} considered what happens when one applies the message passing method to a network that was itself generated from the stochastic block model.  In the simplest case, suppose we generate a network with an even number $n$ of nodes divided into two equally sized groups, and suppose the edge probabilities~$\omega_{rs}$ take just two values, $\oin$~when $r=s$ (within-group edges) and $\oout$ when $r\ne s$ (between-group edges).

Decelle~\etal\ observe that in this situation the message passing equations, Eq.~\eqref{eq:commmp2}, have a symmetric solution $\mu_{i\from j}^r=\frac12$ for all $i,j,r$, which also implies~$q_i^r=\frac12$.  If the message passing iteration converges to this solution then it has failed to find any community structure in the network---every node is identically assigned, half-and-half, to both communities.  Thus the algorithms fails if the symmetric solution is a stable fixed point of the iteration.  Only if it is unstable can we find a nontrivial solution.  As with our other message passing examples, it turns out that there is a bifurcation at which the fixed point changes from stable to unstable.  This bifurcation is driven by changing values of the parameters $\oin$ and~$\oout$.  As the values get close together the empirical distinction between within- and between-group edges becomes smaller and smaller, since the two have almost the same probability, and hence there is less and less signal of the communities embedded in the structure of the network.  Beyond a certain point the signal becomes so weak that the algorithm fails to find anything, which is indicated by the symmetric fixed point becoming stable.

For the simple two-group example considered here, Decelle~\etal~showed that the transition falls at the point where
\begin{equation}
\cin - \cout = \sqrt{2(\cin+\cout)},
\label{eq:phase}
\end{equation}
with $\cin = n\oin$ and $\cout = n\oout$.  When the difference $\cin-\cout$ is smaller than this the algorithm fails to find the communities in the network.  This transition point is known as the \defn{detectability threshold} of the network.

But now we make a crucial observation.  The calculation we have described is based on a fit of the stochastic block model to a network that was itself generated from the same model.  But there is no better method for extracting the parameters of a network (or any data set) than fitting it to the model from which it was truly generated.  This means that no other method for detecting the communities in this network can work better than the one described here.  Hence if this method fails then \emph{all} methods must fail.  So all methods must fail below the detectability threshold.

This is a remarkable result.  The question of whether there are communities in the network and where they are has a well-defined answer in this case: we know the communities in this artificial network because we put them in the network ourselves.  And yet, provably, no algorithm is able to find those communities.  In other words, there are questions about networks for which there exist answers, and yet no method can ever find those answers.  We might imagine, if we are clever enough and work hard enough, that we should be able to answer any question, but the work of Decelle~\etal\ tells us this is not so.  Some questions---even questions with well-defined answers---are unanswerable.

However, the result tells us more than this.  This is not merely a statement about theoretical calculations and algorithms, but also about processes going on in the real world.  Suppose there were some process taking place on a network, such as a social process or a biological process, whose outcome depended on whether there were communities in the network.  Then that process would constitute an algorithm for detecting communities: it would give one outcome if there were communities present and another if there were not.  Below the detectability threshold, however, no such algorithm can exist, which implies in turn that no such real-world process can exist either.

In a way, this is good news.  It tells us that we only care about community structure when we are above the detectability threshold.  If the structure in a network is undetectable then it can have no effect on any real-world outcome and so it doesn't matter whether it is present or not.  Thus we only care about the easy cases of community detection, not the hard ones.

\section{Loopy networks and correlated messages}
\label{sec:loopy}
In our discussion so far we have assumed that the neighbors~$j$ of a node~$i$ are not directly connected to one another by single edges or other short paths, or equivalently that our network contains no short loops.  This assumption was necessary, for instance, to ensure that the probabilities of neighbors belonging to the giant cluster were independent in our percolation example, or that all walks that started along a particular edge returned along the same edge in our graph spectrum example.

A network that contains no loops at all is called a tree, and message passing methods are exact on trees.  Message passing methods are not exact but typically work well on networks that have only long loops but no short ones.  For instance, correlations between percolation probabilities on different nodes typically fall off exponentially with distance through the network, so the presence of long connecting paths introduces only small correlations and correspondingly small errors in our message passing results.

Many real networks, however, including especially social and biological networks, exhibit a high density of short loops, particularly triangles, the shortest loops of all.  On such ``loopy'' networks the message passing methods we have described do not work well, giving poor approximations to the true answers for the problems they are designed to solve~\cite{FM98}.  Indeed in some cases the message passing iteration may fail to converge at all on loopy networks, falling into limit cycles or chaotic trajectories instead.  A number of efforts have been made to remedy these issues, with varying degrees of success~\cite{FM98,YFW03,CC06}.  In this section we describe some recent developments that offer a principled framework for bypassing the problem and creating message passing methods that work on loopy networks~\cite{CN19b,KCN21}.

\begin{figure}
\begin{center}
\includegraphics[width=8cm]{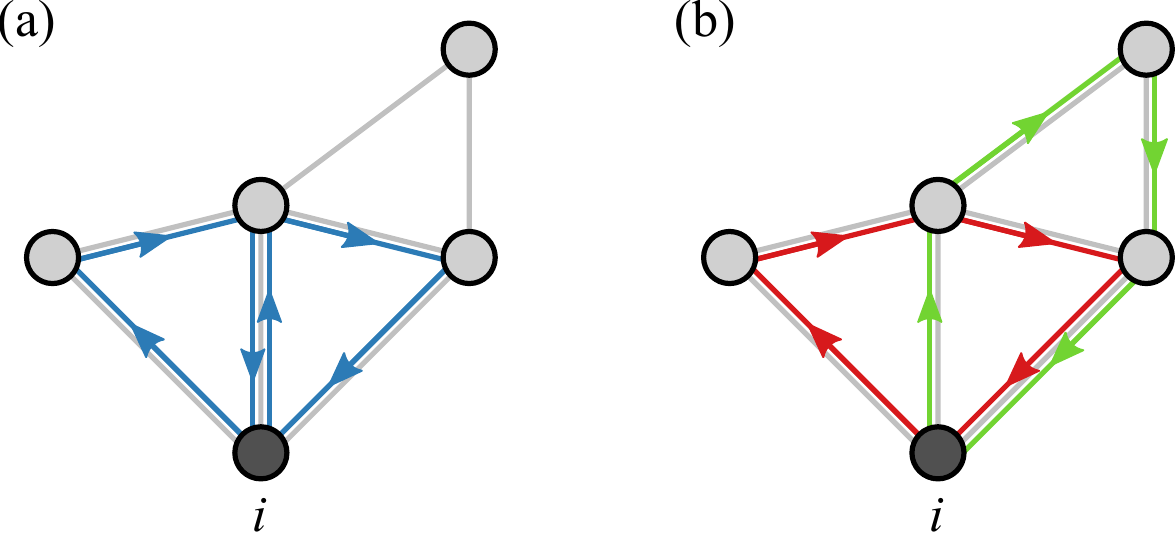}
\end{center}
\caption{(a)~Two cycles of length~3 in a small network, both starting and ending at the same node~$i$.  (b)~Two cycles of length~4 in the same network.  The cycle in green is a primitive cycle but the cycle in red is not because it is composed entirely of edges that were already present in the cycles of length~3.}
\label{fig:loops}
\end{figure}

First, we need to make precise what we mean by a loop.  A \defn{cycle} in a network is a non-self-intersecting walk that starts and ends at the same node.  ``Non-self-intersecting'' here means that no edge is traversed more than once in the walk.  Consider Fig.~\ref{fig:loops}a, which shows the neighborhood of a certain node~$i$ in a network.  Highlighted in blue are two cycles of length~3 starting (and ending) at~$i$.  Figure~\ref{fig:loops}b shows two cycles of length~4 on the same network, highlighted in red and green.  The latter two cycles, however, are qualitatively different from one another in an important way.  The cycle in red follows edges that were already part of the cycles of length~3 in panel~(a).  If we know about the cycles of length~3 then, in a sense, we already know about this cycle of length~4 too.  The cycle in green, however, is new in that it contains some edges we have not seen in any shorter cycle.  We call this a \defn{primitive cycle}.  A primitive cycle from node~$i$ is a cycle starting and ending at~$i$ such that at least one of its edges is not contained in any shorter cycle from node~$i$.  This concept of the primitive cycle is central to understanding how to generalize message passing methods to loopy networks.

Suppose we have a network that contains primitive cycles up to some length~$r$ only and no longer primitive cycles.  It may contain longer non-primitive cycles---potentially many of them---but the longest primitive cycle, from any node anywhere in the network, has length~$r$ or less.  For such a network we can write exact message passing equations according to the following recipe.

\begin{figure}
\begin{center}
\includegraphics[width=4.5cm]{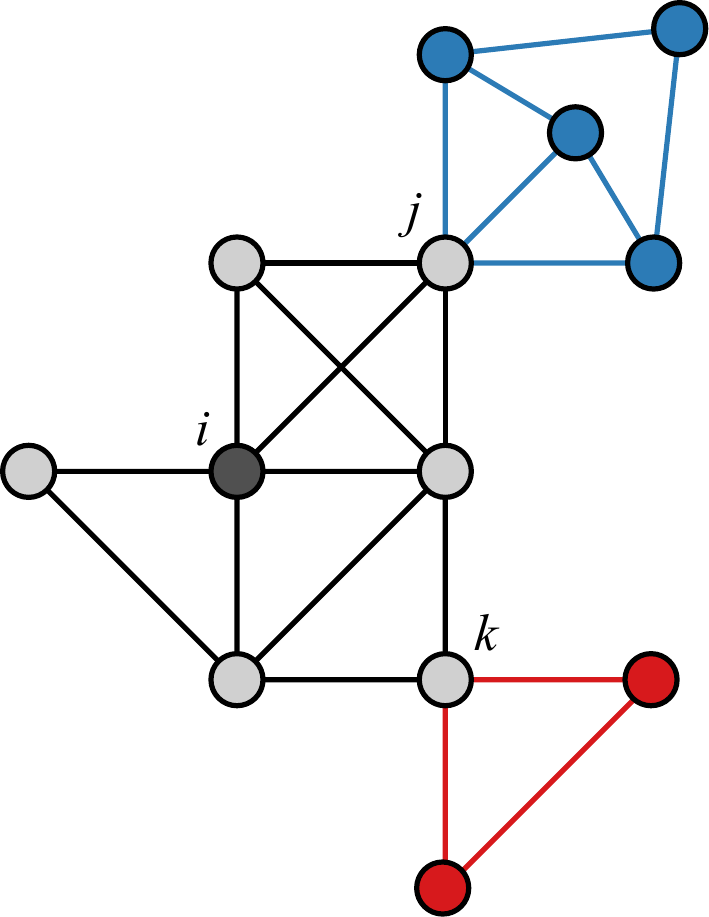}
\end{center}
\caption{The neighborhood~$\mathcal{N}_i$ denoted by the nodes and edges in black and gray contains all primitive cycles of length four or less starting and ending at~$i$.  Node~$i$ receives messages from all other nodes in this neighborhood, which themselves receive messages from their neighborhoods.  Node~$j$, for instance, receives messages from the nodes in blue, but not from the gray nodes of~$\mathcal{N}_i$, which are excluded as described in the text.  We denote the blue neighborhood by~$\mathcal{N}_{i\from j}$.  Note that, according to our definitions, $\mathcal{N}_{i\from j}$~necessarily meets $\mathcal{N}_i$ at only a single node---node~$j$---and also that there cannot be any paths between pairs of neighborhoods such as $\mathcal{N}_{i\from j}$ (in blue) and~$\mathcal{N}_{i\from k}$ (in red), other than through~$\mathcal{N}_i$ itself.  If either of these conditions were violated then there would be primitive cycles of length longer than four starting from~$i$.}
\label{fig:neighborhood}
\end{figure}

Around each node~$i$ in the network we construct a neighborhood~$\mathcal{N}_i$ that consists of the nodes and edges immediately adjacent to~$i$ plus all nodes and edges that lie on primitive cycles starting at~$i$ (of any length up to the network maximum of~$r$).  Figure~\ref{fig:neighborhood} shows an example of such a neighborhood in a network with maximum cycle length~$r=4$.  For a network with $r=2$, which is the smallest possible value, the neighborhoods are the same as in our previous calculations, and the method reduces to standard message passing in this case.  But for $r>2$ the neighborhoods may contain additional edges and nodes.  Note that while, as we have said, there may also be non-primitive cycles starting from~$i$, including ones of length greater than~$r$, these are automatically contained within $i$'s neighborhood, since they are by definition made up entirely of edges that belong to primitive cycles of length~$r$ or less, which are themselves contained within the neighborhood.  Thus under this definition all cycles starting from~$i$, of any length, are contained within the neighborhood of~$i$.

In our extended message passing scheme, node~$i$ receives messages~$\mu_{i\from j}$ from all nodes $j$ in~$\mathcal{N}_i$.  This means that some messages may come from nodes that are not directly connected to $i$ via an edge.  As in normal message passing the messages from neighborhood nodes are themselves calculated from other messages received from outside the neighborhood---see Fig.~\ref{fig:neighborhood} again.  But here a useful simplification occurs.  Since the neighborhood~$i$ includes all cycles of any length that start from node~$i$, it follows that there are no paths connecting the nodes sending messages into the neighborhood, other than through the neighborhood itself.  It there were, they would create cycles outside the neighborhood, of which, by hypothesis, there are none.  This observation is sufficient now to reestablish independence among the incoming messages and allow us to write message passing equations that work on these loopy networks.

This is not quite the whole story, however, for several reasons.  First, as discussed at the start of this section, even in standard message passing we normally allow long loops in the network, since these introduce only small errors.  We can do the same here.  We stipulate that the network should contain no primitive cycles of length greater than~$r$ except ``long'' cycles, meaning ones long enough that the correlations they introduce are small in some sense.  Second, it still remains to specify how we combine the messages coming into a node to calculate the node's properties, and this can be nontrivial.  We will see an example in a moment.  Third, real networks typically do not have a sharp cutoff in the length of their cycles as we have assumed, but instead have primitive cycles of all lengths, up to some (usually large) maximum.  In practice, we approximate such networks by counting loops only up to some chosen length and ignoring longer loops.  This turns out often to be an excellent approximation.  Again we will see an example shortly.

\subsection{Example: Percolation}
\label{sec:loopyperc}
As an application of message passing on loopy networks let us look again at percolation with edge occupation probability~$p$, and specifically at the calculation of the probability that a node in a network belongs to the giant percolation cluster.  The first step in solving this problem is to construct the neighborhood around each node, including all primitive cycles of length~$r$ or less, for some value of $r$ that we choose.  See Fig.~\ref{fig:neighborhood} again for an example.  Now each node~$j$ in the neighborhood of~$i$ transmits a message~$\mu_{i\from j}$ with value equal to the probability that node~$j$ is not connected to the giant cluster when $i$ \emph{and all of its neighborhood~$\mathcal{N}_i$} (except for~$j$ itself) are removed from the network.  As in Section~\ref{sec:mp}, this removal ensures that we ignore any nodes~$j$ that are only connected to the giant cluster via~$\mathcal{N}_i$.

\begin{figure}
\begin{center}
\includegraphics[width=4.5cm]{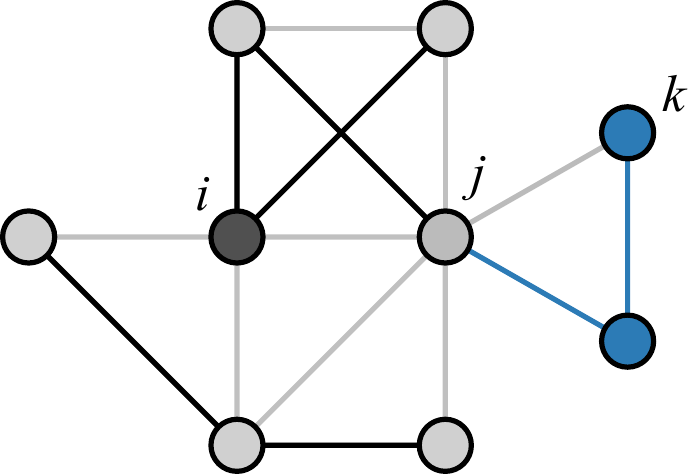}
\end{center}
\caption{The bold edges in this diagram represent ones that are occupied, and for any configuration~$\Gamma_i$ of occupied edges in the neighborhood~$\mathcal{N}_i$ of node~$i$ the probability that $i$ is connected to the giant cluster depends on the probabilities for each of the nodes~$j$ in~$\mathcal{N}_i$ that are reachable along an occupied path.  Similarly, the probability that node~$j$ is connected to the giant cluster in turn depends on the probabilities for each of its reachable neighbors~$k$ in $\mathcal{N}_{i\from j}$.}
\label{fig:percnbh}
\end{figure}

The probability~$\mu_i$ that $i$ itself is not in the giant cluster is now equal to the probability that it is not connected to the giant cluster via any of the nodes $j$ in~$\mathcal{N}_i$.  This probability still requires some effort to calculate.  Figure~\ref{fig:percnbh} shows what is involved.  The edges in the neighborhood~$\mathcal{N}_i$ may be occupied (with probability~$p$) or not (probability~$1-p$) and for every possible configuration~$\Gamma_i$ of occupied edges within the neighborhood let $\sigma_{ij}(\Gamma_i)=1$ if there is at least one path of occupied edges from $i$ to~$j$ within~$\mathcal{N}_i$ and 0 otherwise.  For instance, there is a path from $i$ to~$j$ in Fig.~\ref{fig:percnbh}, so $\sigma_{ij}=1$ for this configuration.

Now the total probability that $i$ is not connected to the giant cluster is given by
\begin{equation}
\mu_i = \sum_{\Gamma_i} P[\Gamma_i] \prod_{j\in\mathcal{N}_i}
  \mu_{i\from j}^{\sigma_{ij}(\Gamma_i)},
\label{eq:loopyperc1}
\end{equation}
with $P[\Gamma_i] = p^m (1-p)^{k-m}$ being the probability of occurrence of the edge configuration~$\Gamma_i$, where $k$ is the number of edges in the neighborhood and $m$ is the number that are occupied in configuration~$\Gamma_i$.  Note how the product in Eq.~\eqref{eq:loopyperc1} computes the total probability that none of the nodes~$j$ to which $i$ is connected belong to the giant cluster and the sum averages this quantity over the probability distribution of configurations~$\Gamma_i$.

There are $2^k$ possible configurations of the $k$ edges in the neighborhood and hence naive evaluation of the sum in~\eqref{eq:loopyperc1} takes time exponential in the neighborhood size.  For small values of the cycle length~$r$ the neighborhoods will also be small, and hence the sum may be tractable.  For larger values it may be necessary to approximate it.  A convenient way of doing this is by Monte Carlo sampling of configurations.  It turns out that one need only sample a small number of configurations to get accurate answers: ten or so is often sufficient.  The reason is that one normally samples similar numbers of configurations for every neighborhood in the network, and the effective number of configurations of the whole network that get sampled is the product of the numbers in each neighborhood.  If one samples ten configurations in the neighborhood of each node then one is effectively sampling $10^n$ configurations of the entire network, which is typically a very large number and more than sufficient to give good convergence of the results.

We still need to calculate the values of the messages themselves, which can be done by a similar method.  Since $\mu_{i\from j}$ depends only on connections outside~$\mathcal{N}_i$, we define a modified neighborhood~$\mathcal{N}_{i\from j}$, which is the normal neighborhood~$\mathcal{N}_j$ with cycles up to length~$r$, minus all nodes and edges in~$\mathcal{N}_i$.  Then
\begin{equation}
\mu_{i\from j} = \sum_{\Gamma_{i\from j}} P[\Gamma_{i\from j}]
  \prod_{k\in\mathcal{N}_{i\from j}} \mu_{j\from k}^{\sigma_{jk}(\Gamma_{i\from j})},
\label{eq:loopyperc2}
\end{equation}
where $\Gamma_{i\from j}$ is a configuration of occupied edges in~$\mathcal{N}_{i\from j}$ and $\sigma_{jk}(\Gamma_{i\from j})=1$ if there is at least one occupied path from $j$ to~$k$ in $\Gamma_{i\from j}$ and 0 otherwise.

Equation~\eqref{eq:loopyperc2} we solve in the standard manner by iteration from any suitable set of starting values, then the converged values are used in~\eqref{eq:loopyperc1} to calculate~$\mu_i$.  From this probability we can also calculate other quantities of interest, such as the expected size~$S$ of the giant cluster, which is given as before by Eq.~\eqref{eq:S}.

\begin{figure}
\begin{center}
\includegraphics[width=\columnwidth]{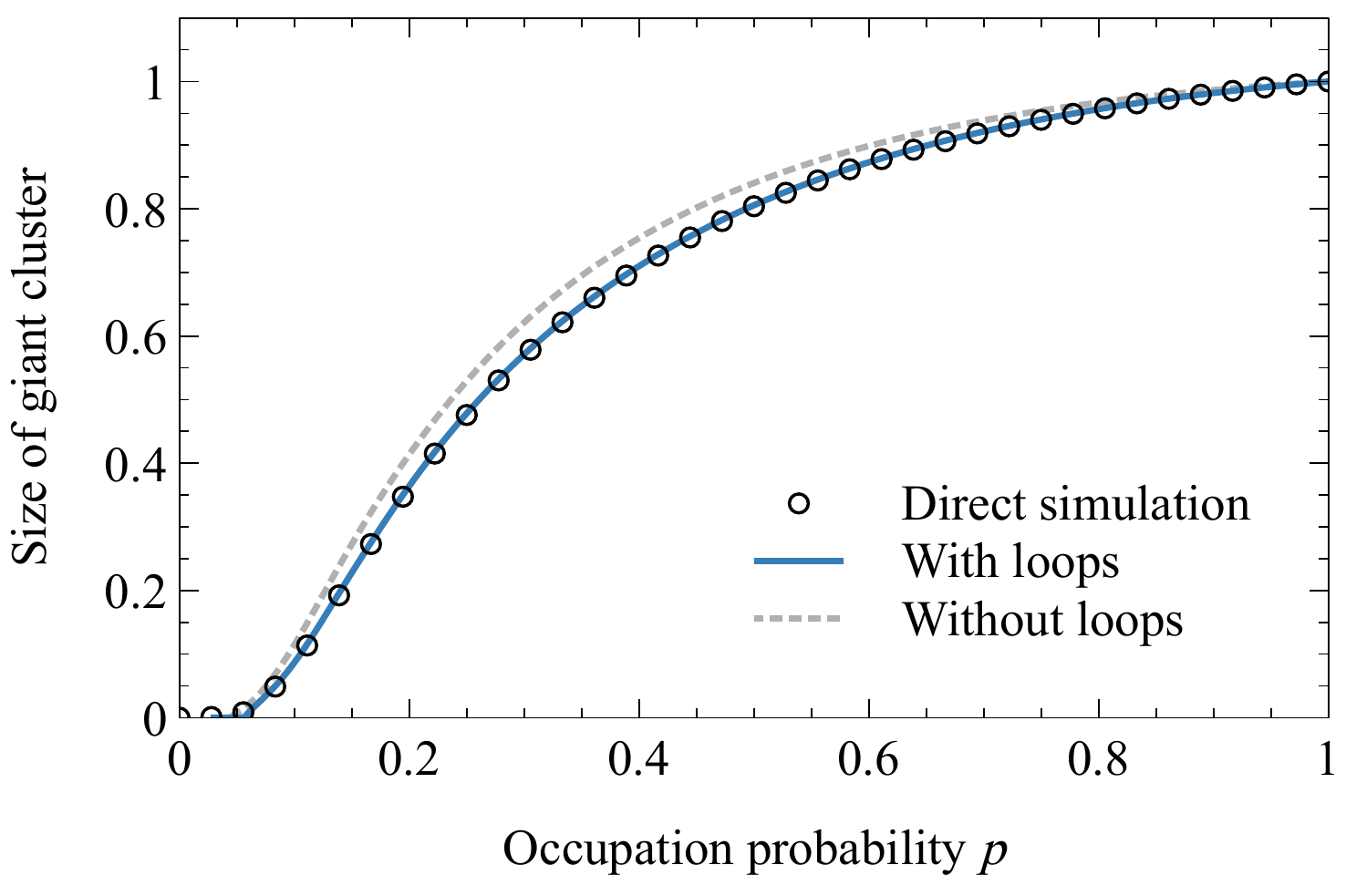}
\end{center}
\caption{Size of the giant cluster for percolation on a large social network of 13\,861 nodes.  The points are results from direct simulation calculations and should be indicative of the ground truth in this case.  The solid curve shows the results of a message passing calculation using Eqs.~\eqref{eq:loopyperc1} and~\eqref{eq:loopyperc2}, including all cycles up to length four.  The dashed line shows what happens when cycles are neglected.  After Cantwell and Newman~\cite{CN19b}.}
\label{fig:loopyperc}
\end{figure}

Figure~\ref{fig:loopyperc} shows an example calculation performed on a real-world network, a social network of 13\,861 nodes with a high density of triangles and other short loops.  The plot shows the size of the giant percolation cluster as a function of the edge occupation probability~$p$, calculated in three ways.  The circles are the results of high-precision direct Monte Carlo simulations (not using message passing), which are laborious to perform but should give an accurate ground-truth result for comparison.  The dashed line is calculated using the standard message-passing method of Eq.~\eqref{eq:perc2}, which does not account for loops, and which does reasonably well in this case but shows clear deviations from the ground truth.  The solid line shows the method of this section, accounting for loops of length up to four, and, as we can see, this calculation now agrees with the ground truth to high accuracy.

It is an open question what the maximum cycle length is that should be incorporated in a calculation like this in order to get good results.  In our work we have found that one can truncate the calculation at surprisingly small cycle lengths---three, four, or five---and still get excellent results, but the exact rate of convergence and the size of the errors introduced are presumably a function of the network structure and the nature of the dependence is not well understood.

\begin{figure}
\begin{center}
\includegraphics[width=\columnwidth]{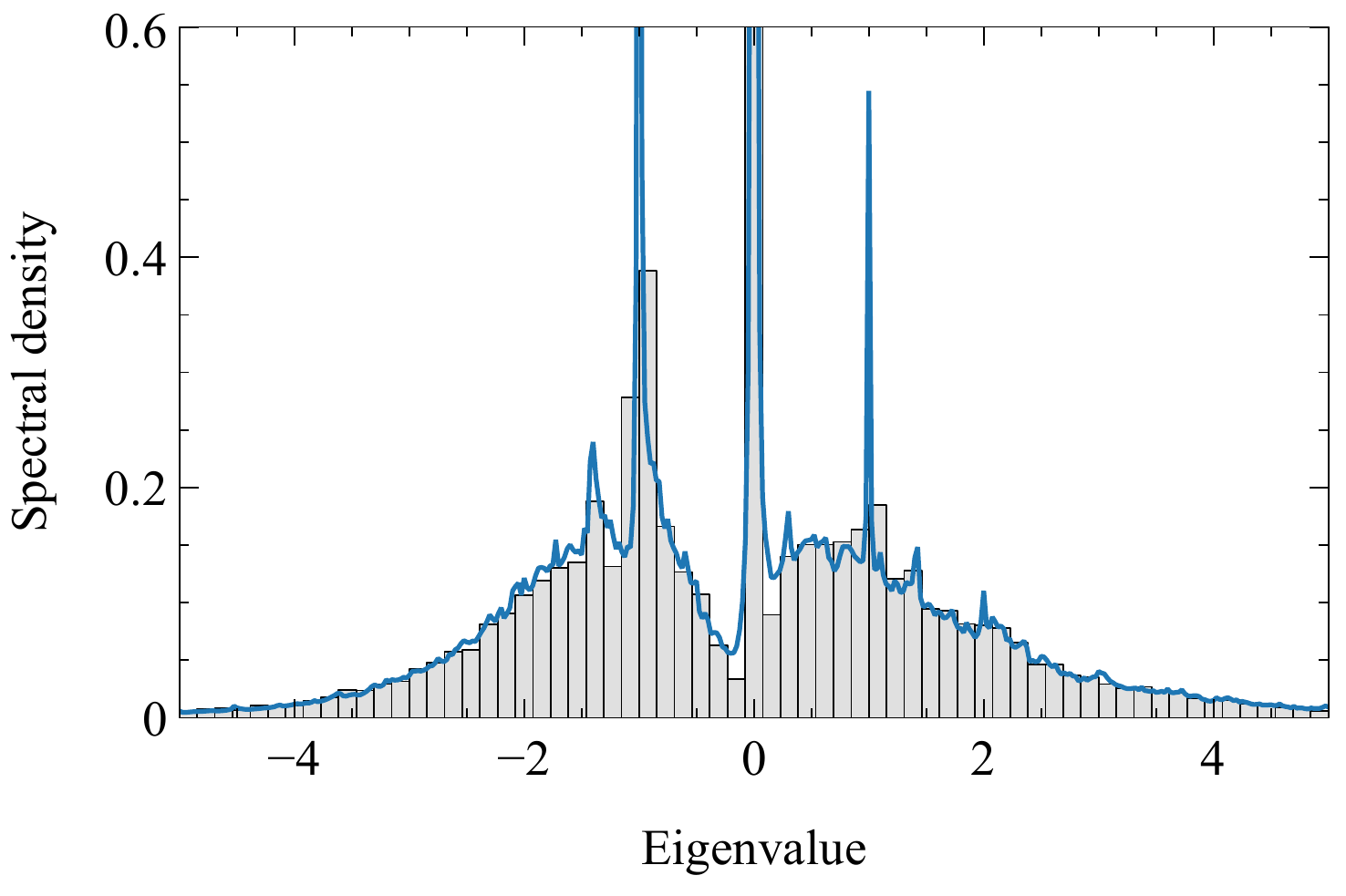}
\end{center}
\caption{The adjacency matrix spectrum of a 10\,680-node PGP network, a combination social/software network of trust relations between PGP keys and their owners.  The histogram shows the distribution of eigenvalues from direct calculations using standard numerical methods, while the curve shows the spectral density from message passing including loops of length up to three.  After Cantwell and Newman~\cite{CN19b}.}
\label{fig:pgp}
\end{figure}

The methods described here can also be extended to other message passing calculations on loopy networks, including the calculation of matrix spectra and the solution of spin models and other probabilistic models~\cite{CN19b,KCN21}.  As an example, Fig.~\ref{fig:pgp} shows a calculation of the spectral density of the adjacency matrix of a large software network---a PGP trust network---and again the message passing calculation compares favorably with the ground truth.  It can also be significantly faster than traditional numerical methods, putting calculations for larger networks within reach: spectra for networks of over 300\,000 nodes have been calculated using these methods on standard (non-parallel) hardware~\cite{CN19b}, something that would be impossible using traditional methods.

\section{Conclusions}
\label{sec:concs}
In this paper we have examined the use of message passing methods for the calculation of node properties on networks.  These methods work by expressing the properties of each node in terms of those of its neighbors, leading to a set of self-consistent equations that are solved by numerical iteration.  This approach has a number of advantages over more conventional numerical methods, being computationally efficient, especially on sparse networks, and allowing one to compute ensemble averages in a single calculation rather than by averaging over repeated simulations.  We have given example applications of message passing to the calculation of percolation properties of networks, the evaluation of graph spectra, community detection in networks and the solution of thermal spin models such as the Ising model.

In addition to its use for numerical methods, we have also described how message passing forms the foundation for further analytic calculations, particularly of phase transition properties.  One can regard the iteration of the message passing equations as a discrete-time dynamical system, whose bifurcations correspond to phase transitions of the original system under study.  Examples include the percolation transition in percolation models, the ferromagnetic transition in the Ising model, and the detectability threshold for community detection.

Finally, we have discussed a shortcoming of the message passing method, at least as it is traditionally formulated, that it works poorly on networks containing short loops because of loss of independence between the states of nearby nodes.  We have reviewed recent work that provides a way around this shortcoming by defining messages that pass not only between immediately adjacent nodes but also between nodes within larger neighborhoods that completely enclose the problematic loops.  This approach leads to more complicated message passing equations but appears to give excellent results in example applications such as percolation and the calculation of matrix spectra.

\hspace{3ex}
\begin{acknowledgments}
The author thanks George Cantwell, Alec Kirkley, Cris Moore, and Austin Polanco for helpful conversations.  This work was funded in part by the US National Science Foundation under grants DMS--1710848 and DMS--2005899.
\end{acknowledgments}

\end{document}